\documentclass{JHEP3}

\usepackage{amsmath}
\usepackage{amssymb}
\usepackage{graphicx}

\newcommand{\AP}{{\alpha^{\prime}}}
\newcommand{\pd}{\partial}

\newcommand{\Dc}{\mathcal{D}}
\newcommand{\diag}{\mathop{\mathrm{diag}}\nolimits}

\newcommand{\jhepre}{\mathop{\mathrm{Re}}\nolimits}
\newcommand{\jhepim}{\mathop{\mathrm{Im}}\nolimits}

\newcommand{\sign}{\mathop{\mathrm{sign}}}

\title{Cosmic acceleration and crossing of $w=-1$ barrier\\
in non-local Cubic Superstring Field Theory model}

\author{Irina Ya. Aref'eva\\
Steklov Mathematical Institute of Russian Academy of Sciences,
Gubkin st., 8, 119991, Moscow, Russia\\
E-mail: \email{arefeva@mi.ras.ru}}

\author{Alexey S. Koshelev\\
Department of Physics, University of Crete, P.O. Box
2208, 71003, Heraklion, Crete, Greece\\
E-mail: \email{koshelev@physics.uoc.gr}}

\abstract{We show  that the late time rolling of the Cubic
Superstring Field Theory (CSSFT) non-local tachyon  in the FRW
Universe leads to a cosmic acceleration with a periodic crossing of
the $w=-1$ barrier. An asymptotic solution for the tachyon and
Hubble parameter by linearizing the non-local equations of motion is
constructed explicitly. For a small Hubble parameter the period of
oscillations is a number entirely defined by the parameters of the
CSSFT action.}


\preprint{\hepth{0605085}}

\begin{document}

\section{Introduction}

The combined analysis of the type Ia supernovae, galaxy clusters
measurements and WMAP (Wilkinson Microwave Anisotropy Probe) data
brings out clearly an evidence of the accelerated expansion of the
Universe~\cite{Perlm}-\cite{Spergel}. The cosmological acceleration
strongly indicates that the present day Universe is dominated by a
smoothly distributed slowly varying Dark Energy (DE). Recent results
of WMAP \cite{Spergel06} together with Ia supernovae data give a
strong support that the present time DE state parameter is close to
$-1$:
\begin{equation*}
w=-0.97^{+0.07}_{-0.09}
\end{equation*}
or without an a priori assumption that the Universe is flat and
together with large-scale structure and supernovae data
$w=-1.06^{+0.13}_{-0.08}$.

From a theoretical point of view the above mentioned domain of $w$
covers three essentially different cases. The first case, $w>-1$, is
achieved in quintessence models~\cite{Wetterich,Peebles} containing
an extra light scalar field which is not in the Standard Model set
of fields~\cite{Okun}. The second case, $w=-1$, is the cosmological
constant~\cite{S-St,Padmanabhan-rev}.  The third case, $w<-1$, is
called a ``phantom'' one and can be realized by a scalar field with
a ghost (phantom) kinetic term. In this case all natural energy
conditions are violated and there are problems of an instability at
classical and quantum levels~\cite{Caldwell,Woodard}.

Since experimental data do not contradict with a possibility $w<-1$
and moreover a direct search strategy to test inequality $w<-1$ has
been proposed~\cite{0312430} a study of such models attracts a lot
of attention. Some projects \cite{Riess06} explore whether $w$
varies with the time or an exact constant. Varying $w$ obviously
corresponds to a dynamical model of DE (see \cite{sami_review} for a
review) which generally speaking includes a scalar
field\footnote{Modified models of GR also generate an effective
scalar field (see for example \cite{modGR} and refs. therein).}.

A possible way to evade the instability problem for models with
$w<-1$ is to yield a phantom model as an effective one, arising from
a more fundamental theory without a negative kinetic term. In this
paper we develop in more details a cosmological SFT tachyon model
\cite{Arefeva}.  The model is based on an SFT formulation of a
fermionic NSR string with the GSO$-$ sector \cite{NPB}. In this
model a scalar field  is the open string tachyon, which describes
according to the Sen's conjecture~\cite{Sen-g} a dynamical
transition of a non-BPS D-brane\footnote{ DE models based on
brane-world scenarios are presented in~\cite{brane}.} to a stable
vacuum (see \cite{review-sft} for review). Since the concerned model
is a string theory limit, all stability issues are related to a
stability of a VSFT (a Vacuum String Field Theory, i.e. the SFT  in
a true vacuum) and one has to discuss only an application of this
limit to a full string theory. There are general arguments \cite{SD}
that there does not exist a local scalar field model for a phantom
Universe without an UV pathology. In a recent paper \cite{VR} it has
been proposed a phantom model without UV pathology in which a vector
field is used.

The scalar model we investigate in this paper is a nonlocal one. Our
goal is a construction of an analytic solution to linearized
Friedmann equations in this model at large times. A characteristic
feature of this model in the flat background is a presence of a
rolling tachyon solution \cite{AJK,yar,VV}. This property persuades
us to consider a fermionic string instead of bosonic one where such
a solution does not exist \cite{Zw,VSV}. However one might expect an
existence of a rolling solution in bosonic strings in a non-flat
case. The dynamics of a non-local tachyon scalar field on a
cosmological background in the Hamilton-Jacobi formalism  is studied
in \cite{calcagni}.

We find that during the late time evolution in the FRW Universe the
tachyon goes to its minimum oscillating with an exponentially
decreasing amplitude. This is similar to the flat case. Consequently
the Hubble parameter goes to a constant, and the state and
deceleration parameters go to $-1$ all oscillating around their
asymptotic values. The DE state parameter $w$ crosses the phantom
divide $w=-1$ during an evolution.

Models with a crossing of the $w=-1$ barrier are also a subject of
recent studies. Simplest models include two fields (one phantom and
one usual field, see \cite{AKVtwofields,Bo,Wei} and refs. therein).
General $\kappa$-essence models~\cite{mukhanov} can have both $w<-1$
and $w\geqslant -1$ but a dynamical transition from the region
$w\geqslant -1$ to the region $w<-1$ or vice versa is forbidden
under general assumptions~\cite{Vikman} and is possible only under
special conditions~\cite{andrianov}.

In our case a non-locality provides a crossing of the $w=-1$ barrier
in spite of the presence of only one scalar field. Hence the
Universe exhibits an acceleration but because of oscillations
quintessence and phantom phases change one each other with time.

The paper is organized as follows. In Section~2 we setup the
cosmological model which is an approximation of the CSSFT describing
a non-BPS brane within the level truncation scheme in the FRW
Universe. In Section~3 we present details of the tachyon dynamics in
the flat case at large times where an approximation linear in
fluctuations around a non-perturbative vacuum is valid. We compare
this linear approximation with a numeric solution to full equations.
In Section~4 we study the tachyon dynamics in the FRW background
again using a linear approximation to the Friedmann equations and
find out that the obtained solution describes an accelerating
Universe. In Section~5 we discuss cosmological consequences of the
obtained results and point out further directions of studying this
type of models.

%%%%%%%%%%%%%%%%%%%%%%%%%%%%%%%%%%%%%%%%%%%%%%%%%%%%%%%%%%%%%%%%%%%%%%%%%%%%%%%%
%%%%%%%%%%%%%%%%%%%%%%%%%%%%%%%%%%%%%%%%%%%%%%%%%%%%%%%%%%%%%%%%%%%%%%%%%%%%%%%%

\section{Setup}

%%%%%%%%%%%%%%%%%%%%%%%%%%%%%%%%%%%%%%%%%%%%%%%%%%%%%%%%%%%%%%%%%%%%%%%%%%%%%%%%

An action for the tachyon in the CSSFT~\cite{AMZ,PTY} in the flat
background\footnote{We always use the signature $(-,+,+,+,\dots)$.}
when fields up to zero mass are taken into account is found to
be~\cite{NPB,AJK}
\begin{equation*}
S_{\text{SFT}}= \frac{1}{g_o^2 \AP ^2} \int dx \left(u^2(x)
-\frac{\AP}2
\eta^{\mu\nu}\pd_{\mu}\phi(x)\pd_{\nu}\phi(x)+\frac{1}{4}\phi^2(x)+
\frac{e^{2\lambda}}{3}\tilde{\phi}^2(x)\tilde{u}(x) \right)
\end{equation*}
where $\phi(x)$ is the tachyon field, $u(x)$ is an auxiliary field,
$$
\tilde{\phi}=e^{\AP\lambda\Box}\phi,
$$
and $\lambda=-\log\frac4{3\sqrt{3}} \approx0.2616$. $\eta$ is the
flat Minkowskian metric, $\Box=\eta^{\mu\nu}\pd_{\mu}\pd_{\nu}$. For
simplicity we will use hereafter $\AP=1$ units in which all fields,
coordinates and the coupling constant $g_o$ are dimensionless.

An auxiliary field $u(x)$  can be integrated out to yield
\begin{equation*}
S_{\text{tach}}=\frac1{g_o^2}\int dx\left(-\frac12
\eta^{\mu\nu}\pd_{\mu}\phi(x)\pd_{\nu}\phi(x)+\frac1{4}\phi^2(x)-
\frac{e^{4\lambda}}{36}\left(\widetilde{\tilde{\phi}^{~2}(x)}\right)^2(x)\right).
%\label{ST0}
\end{equation*}
A reasonable but an ad hoc assumption\footnote{As it was shown in
\cite{AJK,yar} this assumption can be applied to study a special
class of rolling solutions and related questions.} that $u$ has no
the tilde simplifies the last term in this action. Namely, under
this assumption and a rescaling $x\to 2\sqrt{2\lambda}x$, $\phi \to
\frac{3}{\sqrt{2}}e^{-2\lambda}\phi$, and $g_o\to 12\lambda
e^{-2\lambda}g_o$ the action for the tachyon becomes
\begin{equation}
S_{\text{tach, approx}}=\frac1{g_o^2}\int dx\left(-\frac{\xi^2}{2}
\eta^{\mu\nu}\pd_{\mu}\phi(x)\pd_{\nu}\phi(x)+\frac1{2}\phi^2(x)-
\frac{1}{4}(e^{\frac{1}{8}\Box }\phi )^4(x)\right) \label{ST0approx}
\end{equation}
where $\xi^2=\frac1{4\lambda}\approx 0.9556$. The last term in this
action contains an infinite number of derivatives. Just due to this
nonlocal factor a novel behavior in a dynamics of the tachyon field
appears \cite{AJK}.

%%%%%%%%%%%%%%%%%%%%%%%%%%%%%%%%%%%%%%%%%%%%%%%%%%%%%%%%%%%%%%%%%%%%%%%%%%%%%%%%

Cosmological scenarios with our Universe to be considered as a
D3-brane embedded in 10-dimensional space-time was proposed in
\cite{Arefeva}  and a dynamics of this brane is given by the
following covariant version of action (\ref{ST0approx}) in a
non-flat space
\begin{equation}
S=\int
dx\sqrt{-g}\left(\frac{R}{2\kappa^2}+\frac1{g_o^2}\left(-\frac{\xi^2}{2}
g^{\mu\nu}\pd_{\mu}\phi(x)\pd_{\nu}\phi(x)+\frac1{2}\phi^2(x)-
\frac{1}{4}\Phi^4(x)-\Lambda\right)\right)\label{action}
\end{equation}
where
\begin{equation*}
\Phi=e^{\frac18\Box_g}\phi,
\quad\Box_g=\frac1{\sqrt{-g}}\pd_{\mu}\sqrt{-g}g^{\mu\nu}\pd_{\nu}.
\end{equation*}
Here $g$ is the metric, $\kappa$ is  a gravitational coupling
constant and we choose such units that it is dimensionless,
$\Lambda$ is a constant.

In the present analysis we focus on the four dimensional Universe
with the spatially flat FRW metric which can be written as
\begin{equation*}
g_{\mu\nu}=\diag(-1,a^2,a^2,a^2)
\end{equation*}
with $a=a(t)$ being a space homogeneous scale factor. In this
particular case $\Box$ is expressed through $a$ as
\begin{equation*}
\Box_g=-\pd_t^2-3H\pd_t+\frac1{a^2}\pd_{x_i}^2
\end{equation*}
where $H\equiv\dot{a}/a$ is the Hubble parameter and the dot denotes
the time derivative.

The Friedmann equations for the space homogeneous tachyon field
have the form~\cite{Arefeva}
\begin{subequations}
\label{EOM_ST0approx}
\begin{eqnarray}
3H^2&=&\frac{\kappa^2}{g_o^2}\left(\frac{\xi^2}{2}\dot
\phi^2-\frac{1}{2}\phi^2+\frac{1}{4}\Phi^4+{\cal E}_1+{\cal E}_2+\Lambda\right), \label{EOM_ST0approx1}\\
\dot H&=&\frac{\kappa^2}{g_o^2}\left(-\frac{\xi^2}{2}\dot
\phi^2-{\cal E}_2\right) \label{EOM_ST0approx2}
\end{eqnarray}
where
\begin{equation*}
{\cal E}_{1}= -\frac{1}{8} \int_0^1 ds( e^{\frac{1}{8} s {\cal
D}}\Phi^3 ){\cal D} e^{-\frac{1}{8} s {\cal D}} \Phi,\quad {\cal
E}_2= -\frac{1}{8} \int_0^1 ds(
\partial_{t}e^{\frac{1}{8} s {\cal D} }\Phi^3 )\partial_t
e^{-\frac{1}{8}s{\cal D}} \Phi
%\label{nonloc}
\end{equation*}
\end{subequations}
with
\begin{equation*}
\Phi=e^{\frac18 {\cal D}} \phi,\quad {\cal D}=-\pd _t^2-3H(t)\pd_t.
\end{equation*}
The equation of motion for the tachyon is
\begin{eqnarray}
\label{EOM_ST0approx_phi}
 \left(\xi^2{\cal D}+1\right)e^{-\frac{1}{4}{\cal D}}\Phi =\Phi ^3.
\end{eqnarray}
The latter equation is in fact the continuity equation for the
cosmic fluid. To see this explicitly the following operator relation
is useful
\begin{equation*}
\begin{split}
&\lambda\int_0^1ds\left((e^{s\lambda\Box_g}\Box_g\phi)
Pe^{(1-s)\lambda\Box_g}\psi-(e^{s\lambda\Box_g}\phi) \Box_g
Pe^{(1-s)\lambda\Box_g}\psi\right)=\\
=&(e^{\lambda\Box_g}\phi) P\psi-\phi
Pe^{\lambda\Box_g}\psi-\lambda\int_0^1ds(e^{s\lambda\Box_g}\phi) \pi
e^{(1-s)\lambda\Box_g}\psi
\end{split}
\end{equation*}
where $P$ and $\pi$ are operators satisfying $[\Box_g,P]=\pi$.

Equations (\ref{EOM_ST0approx}) are complicated because of the
presence of an infinite number of derivatives and a non-flat metric.
Before we are going to present a construction of an asymptotic
solution to these equations to gain an insight into the problem of
dealing with an infinite number of derivatives we shortly review the
results of an analysis of action (\ref{ST0approx}) in the flat
space~\cite{AJK,yar,VV}.

%%%%%%%%%%%%%%%%%%%%%%%%%%%%%%%%%%%%%%%%%%%%%%%%%%%%%%%%%%%%%%%%%%%%%%%%%%%%%%%%
%%%%%%%%%%%%%%%%%%%%%%%%%%%%%%%%%%%%%%%%%%%%%%%%%%%%%%%%%%%%%%%%%%%%%%%%%%%%%%%%

\section{A Late time Rolling Tachyon in the flat background}

%%%%%%%%%%%%%%%%%%%%%%%%%%%%%%%%%%%%%%%%%%%%%%%%%%%%%%%%%%%%%%%%%%%%%%%%%%%%%%%%

\label{section_ST0_flat}

Approximate action for the tachyon (\ref{ST0approx}) in the flat
background was already studied in \cite{AJK,yar,VV}\footnote{See
also \cite{Zw,VSV} for a cubic action.} and we will present the
results relevant for the sequel. The equation of motion for space
homogeneous configurations of the tachyon field is found to be
\begin{equation}
\left(-\xi^2\partial_t ^2+1\right)e^{\frac{1}{4}\partial_t
^2}\Phi(t)=\Phi(t)^3. \label{EOM_ST0approx_flat}
\end{equation}
where $\Phi=e^{-\frac18\partial_t ^2}\phi$. The operator
$e^{s\partial_t ^2}$ for positive $s$ can be rewritten in an
integral form and this allows further to implement an iteration
procedure which is capable to find a numeric solution to the
problem.

If we are interesting in the Rolling Tachyon solution then we have
in mind the following picture. The tachyon field starts from the
origin (may be with a non-zero velocity), rolls down to the minimum
of the tachyon potential\footnote{The potential is equivalent in
terms of both $\phi$ and $\Phi$ since $\phi=\Phi$ for zero
momentum.} and eventually stops in the minimum. In our notation the
minima are located at $\Phi_0=\pm1$. For $\xi^2\neq 0 $ and
$\xi^2<\xi_{\text{cr}}\approx 1.38$ there are damping fluctuations
near the minimum \cite{yar}. Let us note that in our case $\xi^2<
\xi^2_{\text{cr}}$. To analyze the late time behavior one can
linearize  equation (\ref{EOM_ST0approx_flat}) as
$\Phi=\Phi_0-\delta\Phi$ keeping only liner in $\delta\Phi$ terms. A
substitution yields the following  equation for $\delta\Phi$
\begin{equation}
\left(-\xi^2\partial_t ^2+1\right)e^{\frac{1}{4}\partial_t
^2}\delta\Phi=3 \delta\Phi. \label{EOM_ST0approx_flat_as}
\end{equation}
At this point we want to mention two facts about this equation.
First, if one omits the non-local exponential operator then one
arrives to an equation for the harmonic oscillator
\begin{equation*}
\left(\xi^2\pd_t^2+2\right)\delta\Phi=0
%\label{EOM_ST0approx_flat}
\end{equation*}
which does not give the desired late time behavior. Second, if we
expand the non-local operator and keep only the constant and second
derivative terms the situation depends on the value of $\xi^2$ and
for $\xi^2>\frac14$ (which is our case) remains qualitatively the
same. Using that $e^{\pd_t^2}e^{-m t}=e^{ m^2}e^{-m t}$ one can try
to find a solution to equation (\ref{EOM_ST0approx_flat_as}) of the
form $\delta\Phi(t)=Ce^{-m t}$ where for $m$ we have the following
characteristic equation
\begin{equation}
\left(-\xi^2m^2+1\right)e^{\frac{1}{4}m ^2}=3.
\label{characteristic_m}
\end{equation}
This is exactly equation (3.27) in \cite{Arefeva} with a
redefinition $m_{\text{our}}=im_{\text{\cite{Arefeva}}}$.

Note, that this solution is valid only for late times. One expects
thereby that the non-local operator plays a crucial role in the
asymptotic regime. The latter characteristic equation for $m$ is
solved in Appendix~\ref{appendix}. Here only the results are used. A
general solution for $\delta\Phi$ is an infinite sum
\begin{equation*}
\delta\Phi(t)=\sum_k(A_ke^{-m_k t}+B_ke^{m_kt})
%\label{solution_deltaphi_flat_general}
\end{equation*}
because infinitely many roots $m_k^2$ exist. This reflects the
presence of an infinite number of derivatives. The latter means an
infinite number of initial conditions which however are not
arbitrary (see \cite{Zw} for a discussion on this point). Moreover
in the full an action asymptotic behavior is subject to initial
conditions in the origin. In other words considering the full
non-linear equation and imposing initial conditions in the beginning
of the evolution one arrives to a specific asymptotic configuration.
We assume in our case a solution converging to an asymptotic value
exists. For the definition of $m_k$ given in Appendix a vanishing
condition imposes $B_k=0$ for all $k$ and a reality condition for
$\delta\Phi$ is formulated as $A_k=A_{-1-k}^*$. Therefore, the most
general real vanishing solution to equation
(\ref{EOM_ST0approx_flat_as}) is
\begin{equation}
\delta\Phi(t)=\sum_{k\geq0}(A_ke^{-m_k t}+A_k^*e^{-m_k^*t}).
\label{solution_deltaphi_flat_general_real}
\end{equation}
Hence one yields a real function of time which oscillates with an
exponentially decreasing amplitude. The function $\Phi(t)$ in turn
goes to $1$ oscillating with the decreasing amplitude near the
asymptotic value. Recall that the solution constructed is valid only
for late times. Let us also note that in a UV region, $k\geq 1$, one
has to take into account string modes next to leading ones and this
is beyond of our approximation.

The main (i.e. the most slowly vanishing) contribution in
(\ref{solution_deltaphi_flat_general_real}) is given by $k=0$ and
can be represented as
\begin{equation}
\delta\phi(t)=Ce^{-r t}\sin(\nu t+\varphi)\quad\text{where}\quad
r\approx1.1365,~\nu\approx1.7051 \label{solution_deltaphi_flat_sin}
\end{equation}
where we passed to $\delta\phi=e^{\frac18\pd_t^2}\delta\Phi$. All
other $k$ will give a faster convergence and will play a role of
small corrections. We find that a rather simple choice of constants
$C\approx 1$ and $\varphi\approx 2$ in
(\ref{solution_deltaphi_flat_sin}) matches the rolling solution
presented in Fig.~\ref{figure_flat_match}a. Figures
Fig.~\ref{figure_flat_match}b and Fig.~\ref{figure_flat_match}c
contain corresponding plots.
 \FIGURE{
\includegraphics[width=4cm]{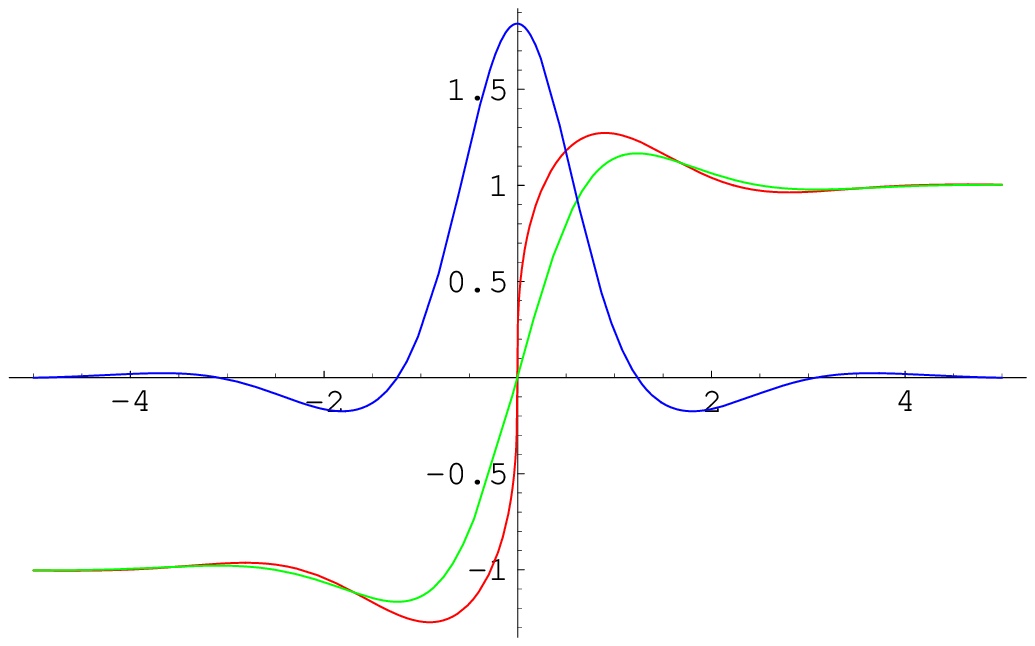}a
\includegraphics[width=4cm]{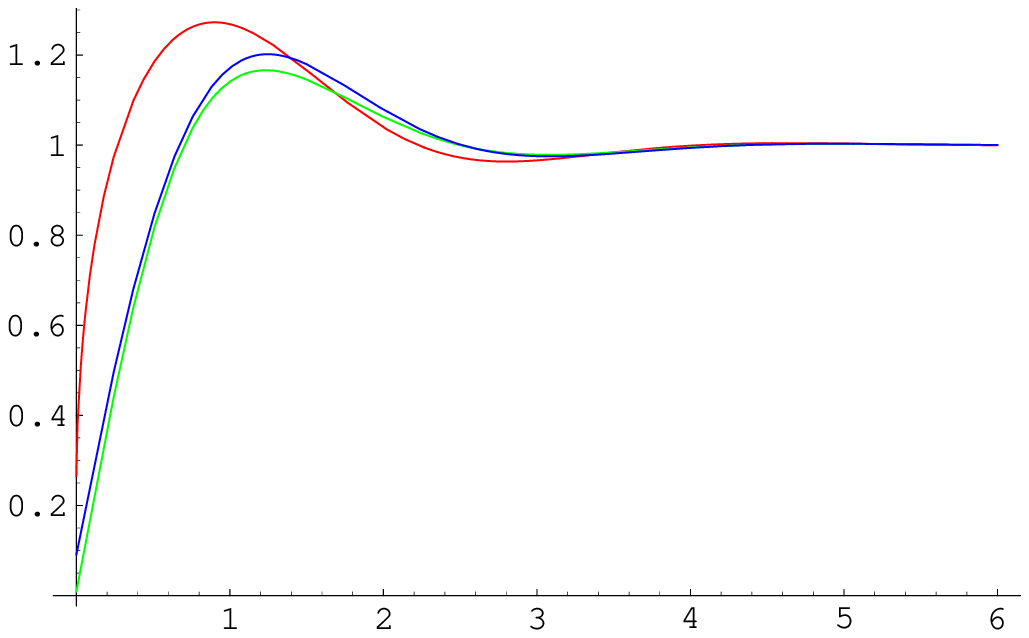}b
\includegraphics[width=4cm]{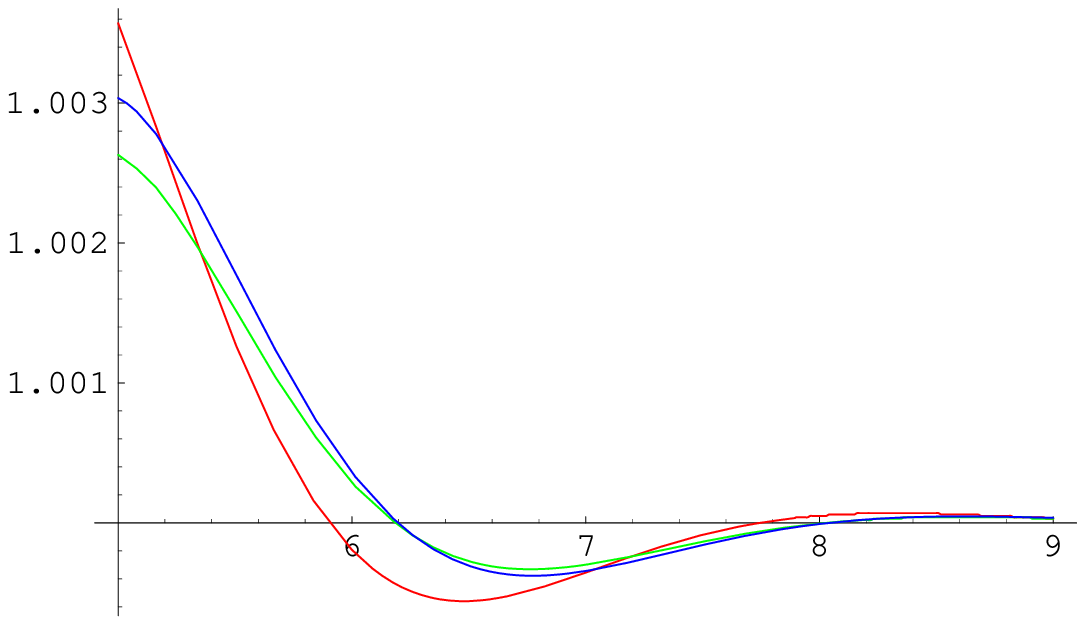}c
\caption{a. Solution to equation~(\ref{EOM_ST0approx_flat}): $\Phi$
(red line), $\phi$ (green line), $\dot\phi$ (blue line) for
$\xi^2=0.9556$; b. An approximation to this solution $1-\delta\phi$
given by (\ref{solution_deltaphi_flat_sin}) (blue line); c. The same
plots as in b in a different scale.} \label{figure_flat_match}}

Note that for $\xi^2=0$ (this case corresponds to a $p$-adic
string~\cite{BFOW,padic} with $p=3$) equation
(\ref{EOM_ST0approx_flat_as}) is simplified drastically and we have
\begin{equation*}
m^2_k=4(\log3+2\pi ki)
\end{equation*}
where again different branches may be considered. The principal
branch is $k=0$ and it corresponds to the rolling solution
\cite{AJK,yar,VV}. Other $k$ do not match the rolling solution. They
will give oscillations which were not present in a numeric analysis.
Absence of such an oscillating behavior is related to initial
conditions in the origin. One can also say that $m_k$ for
$k=1,2,\dots$ are in a UV region beyond of our approximation.

The main conclusion of this Section is that the non-local operator
$e^{s\partial_t ^2}$ for positive $s$ acts like a friction. For
$\xi^2=0$ the late time behavior of the tachyon is just a smooth
rolling with a monotonically vanishing velocity to the minimum of
the potential. This is not surprising mathematically but looks
rather strange physically. Term with $\xi^2$ in the l.h.s. of
equation (\ref{EOM_ST0approx_flat_as}) accelerates the tachyon but
for actual $\xi^2$ a friction becomes stronger and the tachyon
eventually stops in the minimum.

Another point of view is that for the considered $\xi^2$ time
intervals when the kinetic energy on space homogeneous
configurations is negatively defined dominate providing a phantom
behavior for the tachyon field. This justifies a phantom
approximation used in \cite{AKV}.

%%%%%%%%%%%%%%%%%%%%%%%%%%%%%%%%%%%%%%%%%%%%%%%%%%%%%%%%%%%%%%%%%%%%%%%%%%%%%%%%
%%%%%%%%%%%%%%%%%%%%%%%%%%%%%%%%%%%%%%%%%%%%%%%%%%%%%%%%%%%%%%%%%%%%%%%%%%%%%%%%

\section{A Late time Rolling Tachyon in the FRW Universe}
\label{section_ST0_FRW}

We consider action (\ref{action}) for the tachyon coupled to the
gravity. We are going to employ an analog of the asymptotic
expansion used above in Section~\ref{section_ST0_flat}. To this end
we have to expand $\phi=\phi_0-\delta\phi$ (and accordingly
$\Phi=\phi_0-\delta\Phi$, where
$\delta\Phi=e^{\frac18\Dc}\delta\phi$) in Friedmann equations
(\ref{EOM_ST0approx}). In our notations $\phi_0=\pm1$ and the
resulting equations read
\begin{subequations}
\label{EOM_ST0approx_as}
\begin{eqnarray}
3H^2&=&\frac{\kappa^2}{g_o^2}\left(\frac{\xi^2}2\dot{\delta\phi}^2-\frac12\delta\phi^2+\frac32\delta\Phi^2+\delta{\cal E}_1+\delta{\cal E}_2+\Lambda_0\right), \label{EOM_ST0approx_as1}\\
\dot
H&=&\frac{\kappa^2}{g_o^2}\left(-\frac{\xi^2}2\dot{\delta\phi}^2-\delta{\cal
E}_2\right) \label{EOM_ST0approx_as2}
\end{eqnarray}
where
\begin{equation*}
\delta{\cal E}_1=-\frac38\int_0^1dse^{\frac18s\Dc}\delta\Phi\Dc
e^{-\frac18s\Dc}\delta\Phi,\quad \delta{\cal
E}_2=-\frac38\int_0^1ds\pd_te^{\frac18s\Dc}\delta\Phi
\pd_te^{-\frac18s\Dc}\delta\Phi\nonumber
\end{equation*}
\end{subequations}
and $\Lambda_0=-\frac14+\Lambda$. The equation of motion for the
tachyon field $\phi$ (\ref{EOM_ST0approx_phi}) after expansion has
the form
\begin{equation}
(\xi^2\Dc+1)e^{-\frac14\Dc}\delta\Phi=3\delta\Phi.
\label{EOM_ST0approx_as_phi}
\end{equation}
At this point we assume that in this approximation only the constant
term $H_0$ in an expansion of $H$ survives in the latter equation. A
validity of this assumption will become clear after a solution will
be constructed. A value of $H_0$ can be determined from equation
(\ref{EOM_ST0approx_as1}). However, even in this case an analytic
solution to equation (\ref{EOM_ST0approx_as_phi}) in a closed form
is not achievable so far. Instead of this we are looking for a
solution constructed from eigenfunctions of the operator
$\Dc_0=-\pd_t^2-3H_0\pd_t$. They have the following form
\begin{equation*}
\delta\Phi=Ae^{-\bar{m}_+ t}+ Be^{-\bar{m}_-
t}\quad\text{where}\quad
\bar{m}_{\pm}=\frac32H_0\pm\sqrt{\frac{9H_0^2}4+m^2}.
%\label{solution_deltaphi}
\end{equation*}
To become a solution to equation (\ref{EOM_ST0approx_as_phi}) with a
constant $H=H_0$ a parameter $m$ should be determined by means of
transcendental equation (\ref{characteristic_m}) already appeared in
the flat case. A general solution to equation
(\ref{EOM_ST0approx_as_phi}) is
\begin{equation*}
\delta\Phi=\sum_k\left(A_{k}e^{-\bar{m}_{+k}t}+
B_{k}e^{-\bar{m}_{-k}t}\right).
%\label{solution_deltaphi_FRW}
\end{equation*}
An analysis of this solution goes in an analogy with the flat case
considered in Section~\ref{section_ST0_flat}.

For $\bar{m}_{\pm k}$ we have
\begin{equation*}
\begin{split}
&\bar{m}_{\pm k}=\bar{r}_{\pm k}+i\bar{\nu}_k,~ \bar{r}_{\pm
k}=\frac32H_0\pm\frac{|\beta_k|}{\bar{v}},~\bar{\nu}_k=\sign(\beta_k)\frac
{\bar{v}}2,\\
&\bar{v}=\sqrt{-2\left(\frac{9H_0^2}4+\alpha_k\right)+2\sqrt{\left(\frac{9H_0^2}4+\alpha_k\right)^2+\beta_k^2}}
\end{split}
\end{equation*}
where a notation $m_k^2=\alpha_k+i\beta_k$ is used. However, in
contrary with the flat case a selection of vanishing branches is not
so obvious. It depends on a particular value of $H_0$. Notice, that
$\bar{\nu}_k$ is always non-zero, i.e. a solution is oscillating for
any $k$. Fortunately, the symmetry $\bar{m}_{\pm
k}=\bar{m}_{\pm(-1-k)}^*$ persists. Thus a reality condition for
$\delta\Phi$ is $A_k=A_{-1-k}^*$ and $B_k=B_{-1-k}^*$, and  the most
general real vanishing solution becomes
\begin{equation}
\delta\Phi=\sum_{k\geq0}{}^{\prime}\left(A_{k}e^{-\bar{m}_{+k}t}+
A_{k}^*e^{-\bar{m}_{+k}^*t}\right)+\sum_{k\geq0}{}^{\prime}\left(B_{k}e^{-\bar{m}_{-k}t}+
B_{k}^*e^{-\bar{m}_{-k}^*t}\right)
\label{solution_deltaphi_FRW_general_real}
\end{equation}
where the prime means that only $k$ for which the behavior is
vanishing should be taken into account.

Let us spend few lines discussing a question about the main (i.e.
most slowly vanishing) contribution. A parameter $H_0$ complicates
the story since $H_0>0$. Thus $k=0$ is not necessary the main
contribution. Moreover, for specific values of $H_0$ and $k$ both
plus and minus in $\bar{r}_k$ may give a positive number. In this
case a corresponding $B_k$ term is allowed and this is new compared
to the flat case. To illustrate the situation $\bar{m}_k$ are drawn
in Fig.~\ref{figure_roots} for different values of $H_0$.
 \FIGURE{
\includegraphics[width=4cm]{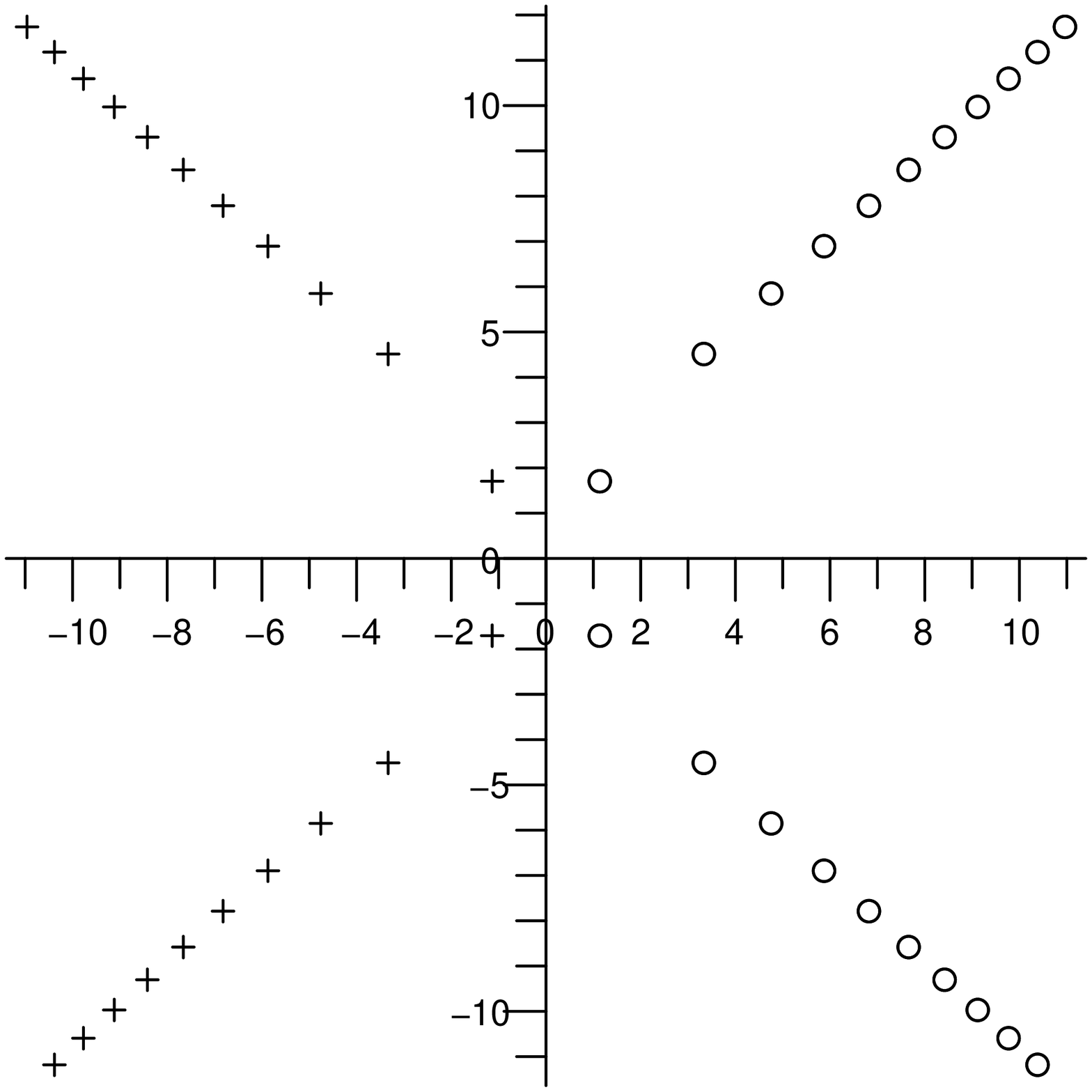}
\includegraphics[width=4cm]{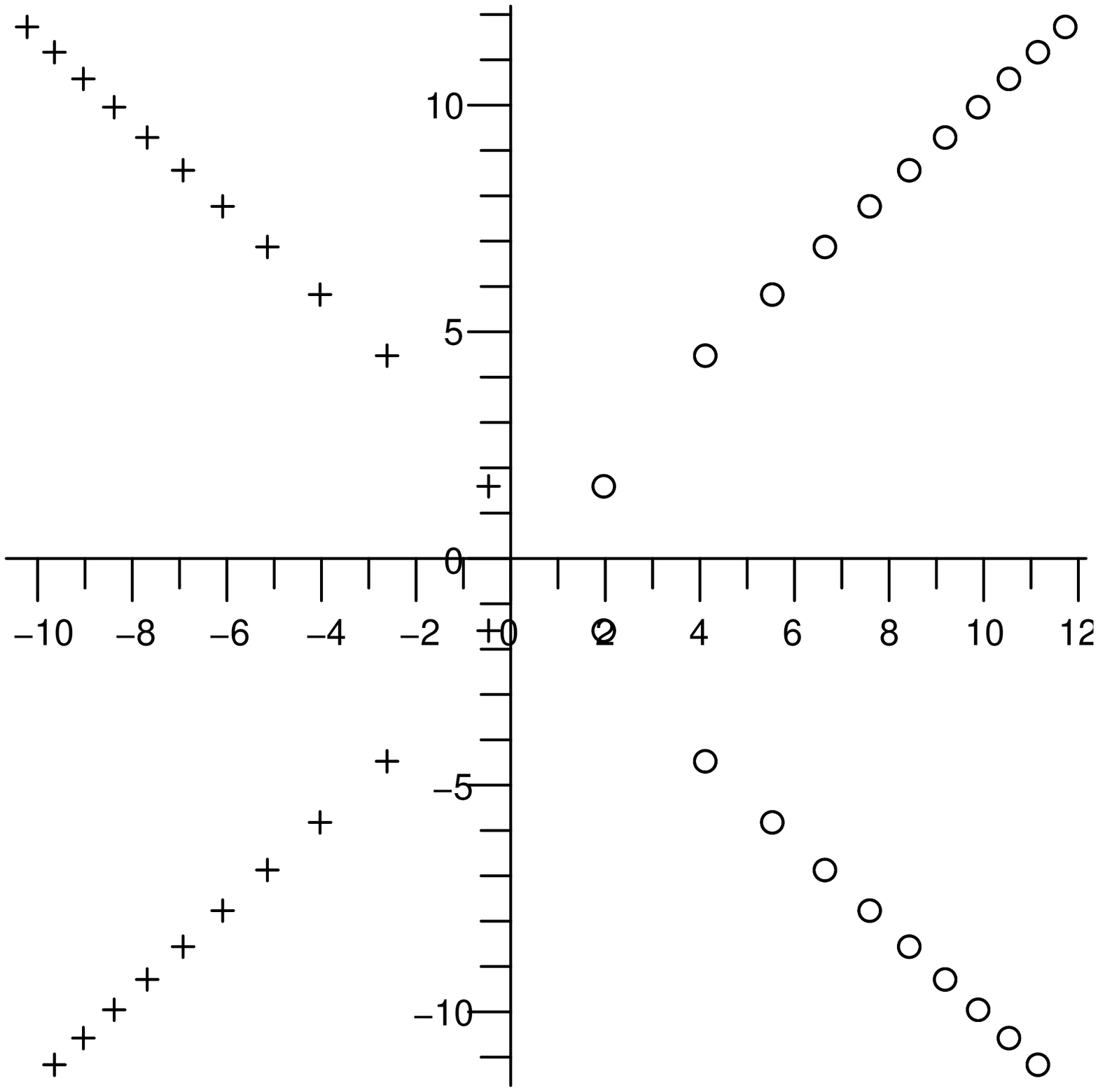}
\includegraphics[width=4cm]{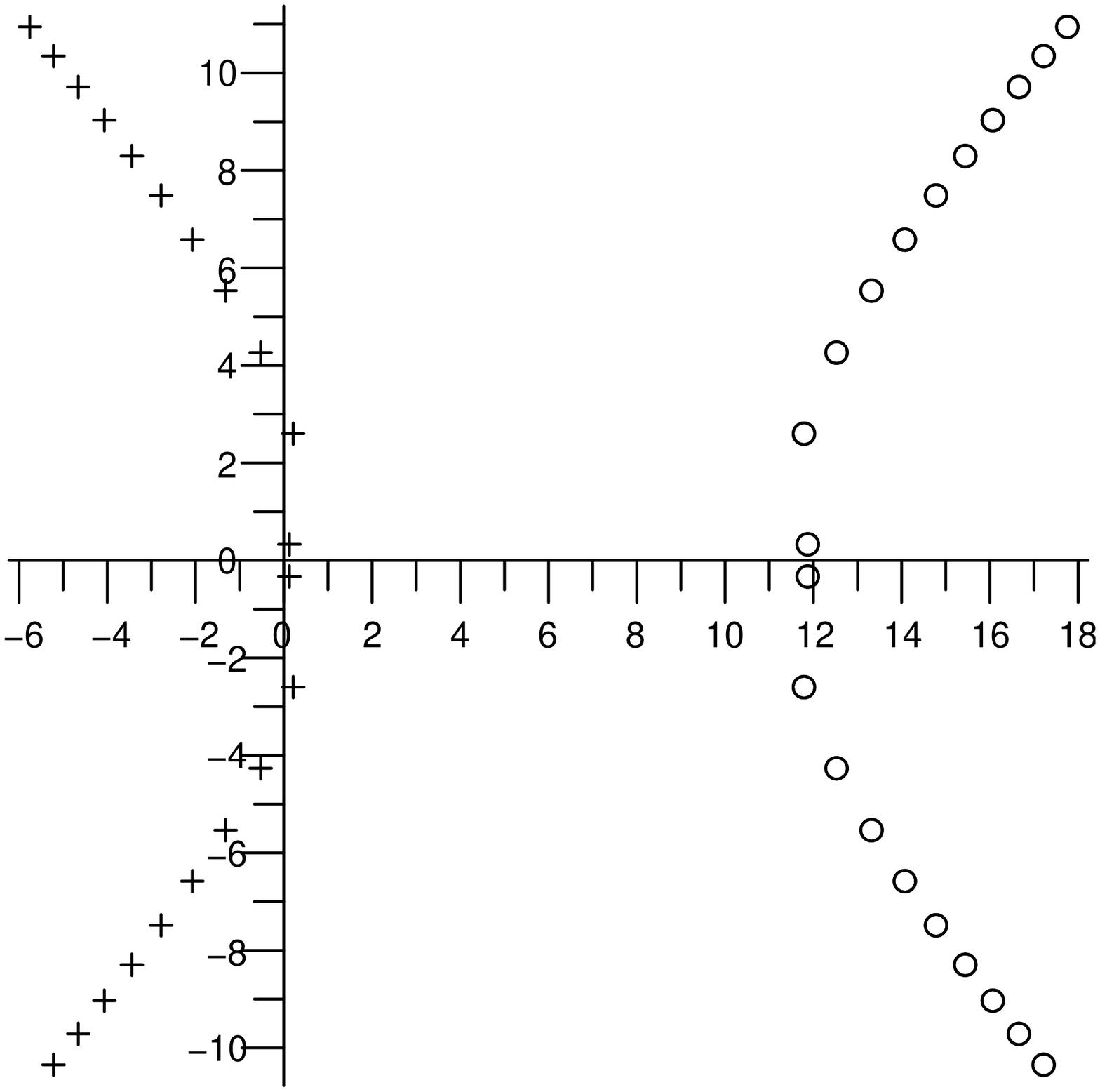}
\caption{$\bar{m}_{\pm k}$ for $H_0=0$ (left), $H_0=\frac12$
(middle), and $H_0=4$ (right). Circles correspond to the plus and
crosses to the minus sign.} \label{figure_roots}}
 On these plots
each point corresponds to a specific $k$. For $k\geq0$ the imaginary
part is positive and grows with $k$. For $k\leq-1$ the imaginary
part is negative and grows in absolute value with $|k|$. Circles
correspond to the plus and crosses to the minus sign in
$\bar{m}_{\pm k}$. Points with smallest positive real parts form the
main contribution and one sees that a selection of such points is
not so transparent. However, for the sequel we will need only to
state that one can make such a selection. Thus for the main
contribution in (\ref{solution_deltaphi_FRW_general_real}) we have
\begin{equation}
\delta\phi=Ce^{-\bar{m}t}+ C^*e^{-\bar{m}^*t}
\label{solution_deltaphi_FRW}
\end{equation}
where we have passed to $\delta\phi=e^{-\frac18\Dc}\delta\Phi$ and
an index $\pm k$ in $m_{\pm k}$ is omitted since a specific choice
of $k$ and a branch of the square root is made. Also for some values
of $H_0$ it is possible that two close points, say $k$ and $k+1$ for
$k>0$ will have equal and smallest positive real parts. In this case
they will only differ by a frequency of oscillations which will be
suppressed with an equal exponential factor. However, for simplicity
we will not consider this situation. Substituting solution
(\ref{solution_deltaphi_FRW}) into Friedmann equation
(\ref{EOM_ST0approx_as2}) one yields after an integration using
relation (\ref{characteristic_m})
\begin{equation}
H=H_0+
\frac{\kappa^2}{g_o^2}\left(\frac{C^2\bar{m}}{16}e^{-2\bar{m}t}\left(1+4\xi^2-\xi^2m^2\right)+
\frac{{C^*}^2\bar{m}^*}{16}e^{-2\bar{m}^*t}\left(1+4\xi^2-\xi^2{m^*}^2\right)\right)
\label{solution_H_FRW}
\end{equation}
where an integration constant is chosen according to equation
(\ref{EOM_ST0approx_as1}) to be
$H_0=\frac{\kappa}{g_o}\sqrt{\frac{\Lambda_0}{3}}$ and a constant
$C$ is the same as in (\ref{solution_deltaphi_FRW}). Further one can
check that equation (\ref{EOM_ST0approx_as1}) is satisfied up to
$\delta\phi^4$ terms which are beyond of our approximation. For the
succeeding analysis it is convenient to represent $\phi$ and $H$ as
follows
\begin{equation}
\phi=1-\bar{C}e^{-\bar{r}t}\sin(\bar{\nu}t+\bar{\varphi})\quad\text{and}\quad
H=\frac{\kappa}{g_o}\sqrt{\frac{\Lambda_0}{3}}-
\frac{\kappa^2}{g_o^2}C_He^{-2\bar{r}t}\sin(2\bar{\nu}t+\varphi_H)
\label{solution_dphH_FRW_sin}
\end{equation}
where $\bar{C}$ and $\bar{\varphi}$ are arbitrary (integration)
constants,
\begin{equation*}
\begin{split}
C_H&=\frac{\bar{C}^2}{32}\sqrt{\bar{m}\bar{m}^*}\sqrt{(1+4\xi^2)^2+\xi^4m^2{m^*}^2-\xi^2(1+4\xi^2)(m^2+{m^*}^2)},\\
\text{and}~\varphi_{H}&=2\bar{\varphi}+\frac{\pi}2+\arctan\left(i\frac{\bar{m}-\bar{m}^*}{\bar{m}+\bar{m}^*}\right)
-\arctan\left(i\frac{\xi^2(m^2-{m^*}^2)}{2+8\xi^2-\xi^2(m^2+{m^*}^2)}\right).
\end{split}
\end{equation*}

In a special case $\xi^2=0$ which corresponds to a $p$-adic string
in the FRW Universe the story is much simpler because eigenvalues
$\bar{m}_{\pm k}$ can be expressed in elementary functions since
$m_k^2=4(\log3+2\pi ki)$. Moreover, $\bar{m}_{\pm0}$ are pure real
and $\bar{m}_{+0}$ corresponds to a smooth approach of the tachyon
field $\phi$ to the asymptotic value $\phi_0=1$ without
oscillations. Expressions for $\delta\phi$ and $H$ in this case are
as follows
\begin{equation}
\delta\phi=\bar{C}e^{-\bar{m}t}~\text{and}~
H=\frac{\kappa}{g_o}\sqrt{\frac{\Lambda_0}{3}}-
\frac{\kappa^2}{g_o^2}C_He^{-2\bar{m}t}
\label{solution_dphH_FRW_q20}
\end{equation}
where $\bar{m}=\frac32H_0+\sqrt{\frac{9H_0^2}4+4\log3}$,
$C_H=\frac{\bar{C}^2\bar{m}}{16}$ and $H_0$ is the same as in
(\ref{solution_H_FRW}). Again one can check that equation
(\ref{EOM_ST0approx_as1}) is satisfied up to $\delta\phi^4$ terms.

Plots representing the tachyon field evolution both for
$\xi^2\approx0.9556$ and $\xi^2=0$ are shown in
Fig.~\ref{figure_phi}.
 \FIGURE{
\includegraphics[width=4cm]{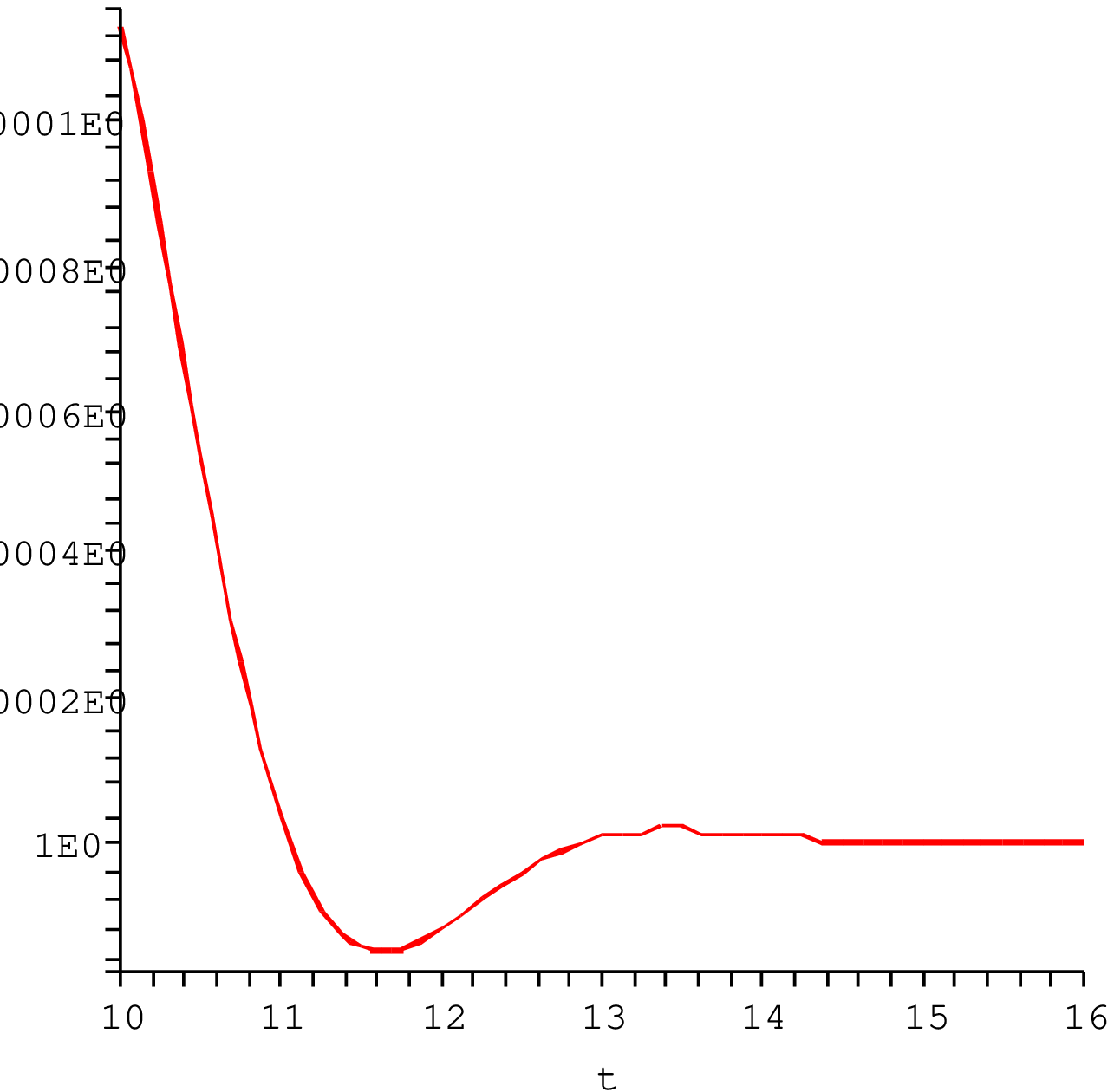}
\includegraphics[width=4cm]{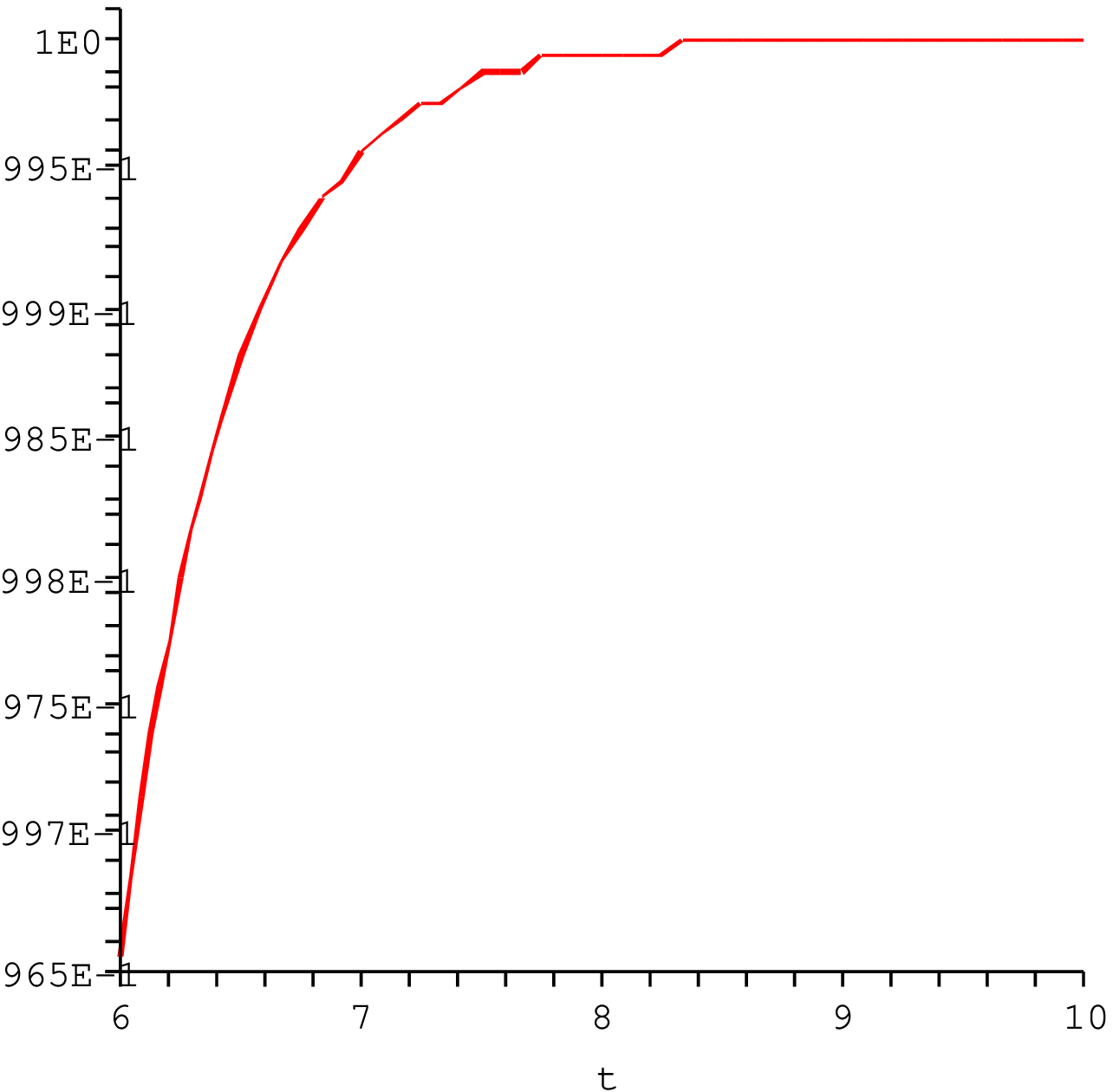}
\caption{Tachyon field for $\xi^2\approx0.9556$ (left) and $\xi^2=0$
(right) for $\bar{C}=1$, $\bar{\varphi}=0$ and
$\Lambda_0=3\cdot10^{-4}$.} \label{figure_phi}}

%%%%%%%%%%%%%%%%%%%%%%%%%%%%%%%%%%%%%%%%%%%%%%%%%%%%%%%%%%%%%%%%%%%%%%%%%%%%%%%%
%%%%%%%%%%%%%%%%%%%%%%%%%%%%%%%%%%%%%%%%%%%%%%%%%%%%%%%%%%%%%%%%%%%%%%%%%%%%%%%%

\section{Cosmological consequences and further directions}

The obtained asymptotic solutions for the tachyon and Hubble
parameter (\ref{solution_dphH_FRW_sin}) lead to a number of
interesting cosmological properties.

First, we mention that $H$ tells us about the time point $t_0$ after
which the solution is reasonable. Indeed, for $H$ to be strictly
positive we have to require $\sqrt{\frac{\Lambda_0}{3}}>
\frac{\kappa}{g_o}C_He^{-2\bar{r}t}$. Solving this inequality w.r.t.
the time we have $t>t_0=-\frac1{2\bar{r}}\log\left(\frac{g_o}{\kappa
C_H}\sqrt{\frac{\Lambda_0}3}\right)$. This is the characteristic
time in question.

Then one sees that during the late time evolution of the tachyon
field the Hubble parameter goes to a constant. $\dot{H}$ obviously
vanishes and moreover, both $H$ and $\dot{H}$ do oscillate. The
state parameter $w$ also has an oscillating time behavior and goes
asymptotically to $-1$. An explicit expression for it is of the form
\begin{equation}
w=-1-\frac{2}{3}\frac{\dot
H}{H^2}=-1+\frac23C_He^{-2\bar{r}t}\frac{-2\bar{r}\sin(2\bar{\nu}t+\varphi_H)+2\bar{\nu}\cos(2\bar{\nu}t+\varphi_H)}
{\left(\sqrt{\frac{\Lambda_0}{3}}-
\frac{\kappa}{g_o}C_He^{-2\bar{r}t}\sin(2\bar{\nu}t+\varphi_H)\right)^2}.
\label{w}
\end{equation}
It is very interesting that $w$ crosses the phantom divide $w=-1$
during the evolution. Such a crossing is forbidden in single field
cosmological models with a local action. In our model we see that a
non-locality breaks this restriction. The deceleration parameter $q$
behaves very similar to $w$ because its expression through $H$ is
very close to $w$
\begin{equation}
q=-1-\frac{\dot
H}{H^2}=-1+C_He^{-2\bar{r}t}\frac{-2\bar{r}\sin(2\bar{\nu}t+\varphi_H)+2\bar{\nu}\cos(2\bar{\nu}t+\varphi_H)}
{\left(\sqrt{\frac{\Lambda_0}{3}}-
\frac{\kappa}{g_o}C_He^{-2\bar{r}t}\sin(2\bar{\nu}t+\varphi_H)\right)^2}.
\label{q}
\end{equation}
Hence the Universe exhibits an acceleration but because of
oscillations quintessence and phantom phases change one each other
with the time.

The scale factor $a$ is related to $H$ as $H=\dot{a}/a$ and can be
readily found to be
\begin{equation}
a=a_0e^{\int
Hdt}=a_0\exp\left(\frac{\kappa}{g_o}\sqrt{\frac{\Lambda_0}{3}}t+
\frac{\kappa^2}{g_o^2}C_He^{-2\bar{r}t}\frac{\bar{r}\sin(2\bar{\nu}t+\varphi_H)+\bar{\nu}\cos(2\bar{\nu}t+\varphi_H)}{\bar{r}^2+\bar{\nu}^2}\right)
\label{a}
\end{equation}
where $a_0$ is an integration constant. For late times it has a
simple exponential approximation
\begin{equation*}
a=a_0e^{\frac{\kappa}{g_o}\sqrt{\frac{\Lambda_0}{3}}t}.
\end{equation*}

To plot $H$ and $w$ we specify $H_0$ (or $\Lambda_0$) to be so small
that the $\bar{r}$ and $\bar{\nu}$ in (\ref{solution_dphH_FRW_sin})
are $\bar{r}_{+0}$ and $\bar{\nu}_{+0}$ respectively. This
corresponds to the middle plot in Fig.~{\ref{figure_roots}}. The
explained above behavior of $H$, $w$, $q$, and $a$ is visualized in
Fig.~\ref{figure_FRW}.
 \FIGURE{
\includegraphics[width=3.5cm]{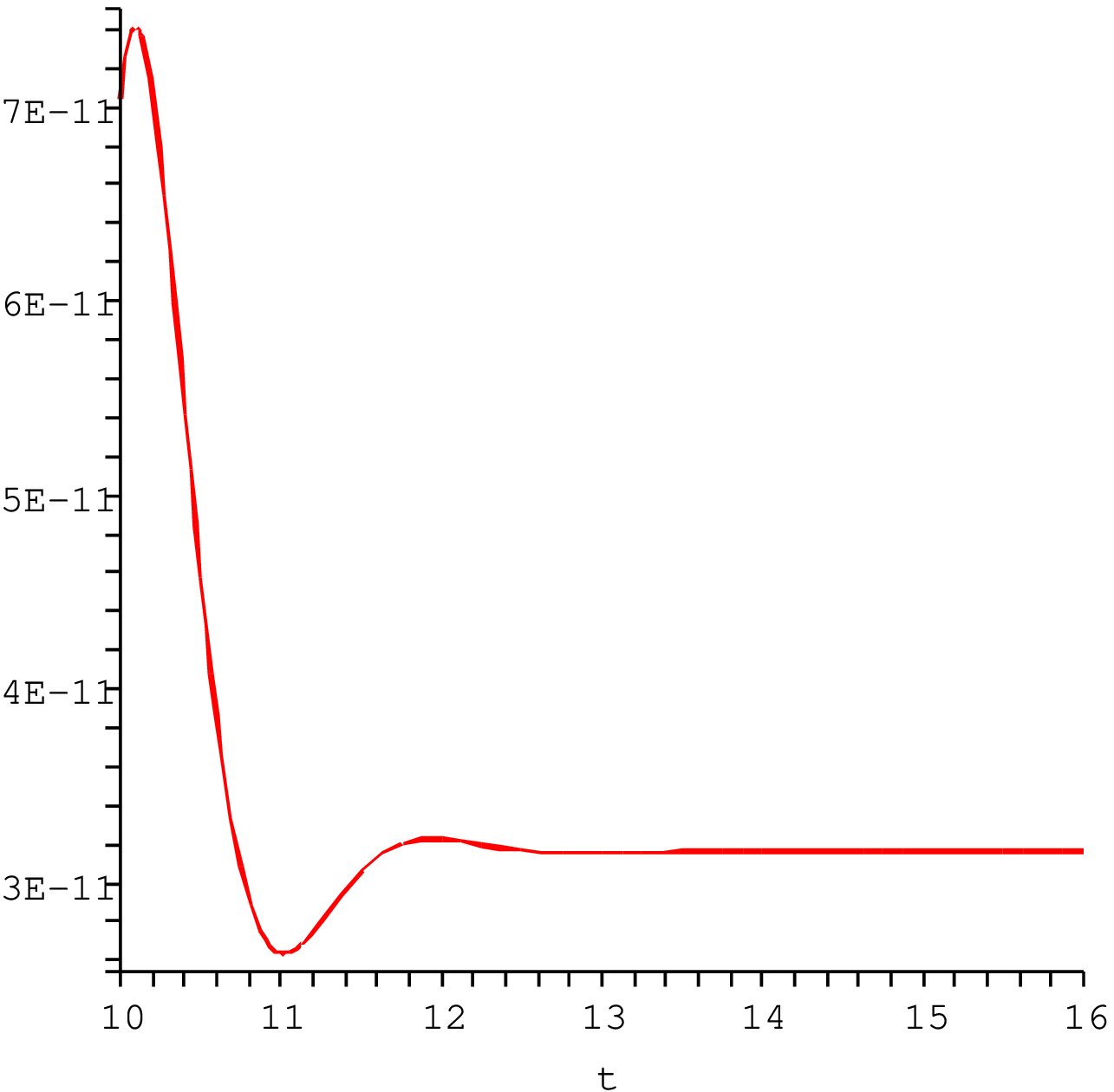}
\includegraphics[width=3.5cm]{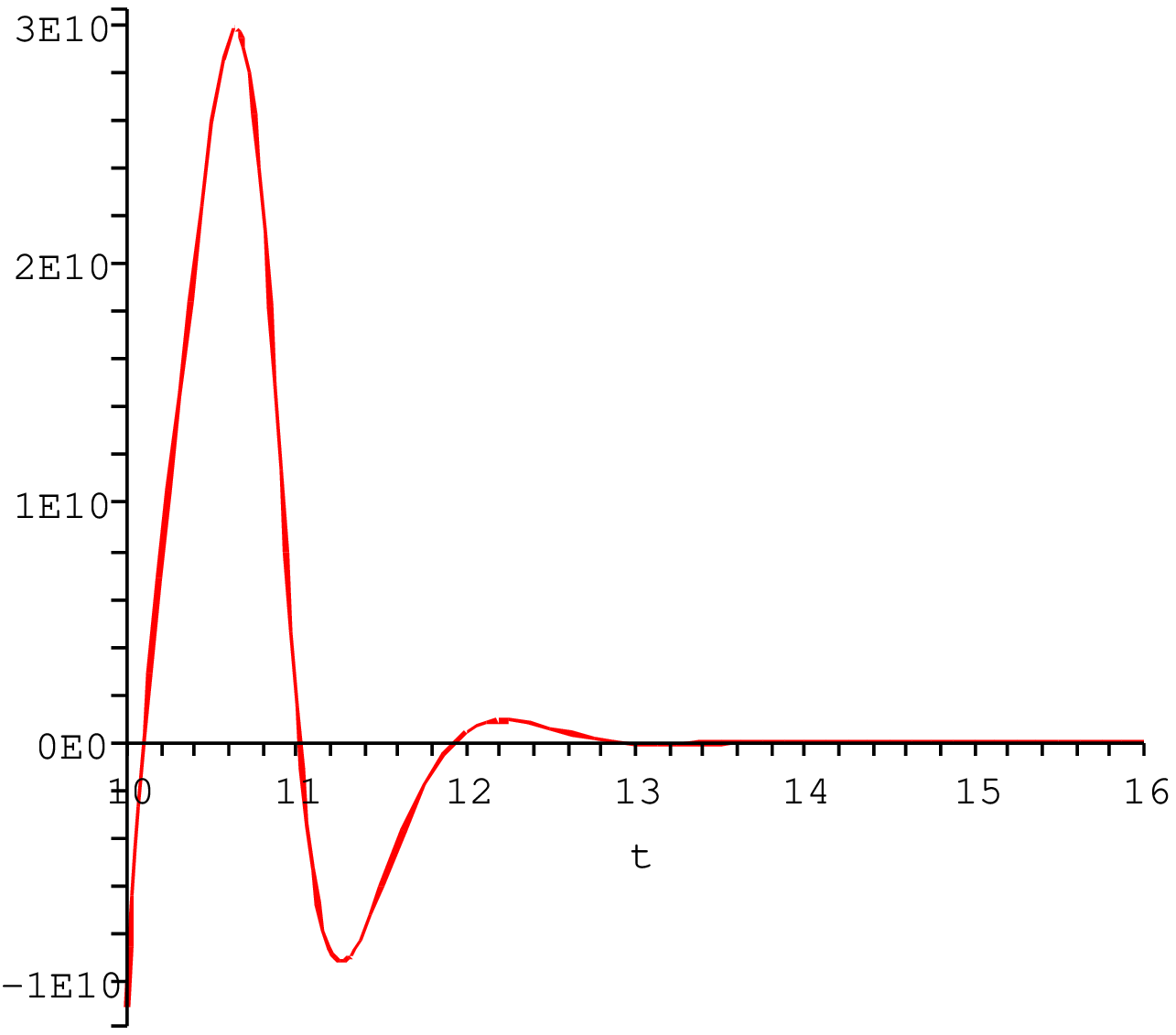}
\includegraphics[width=3.5cm]{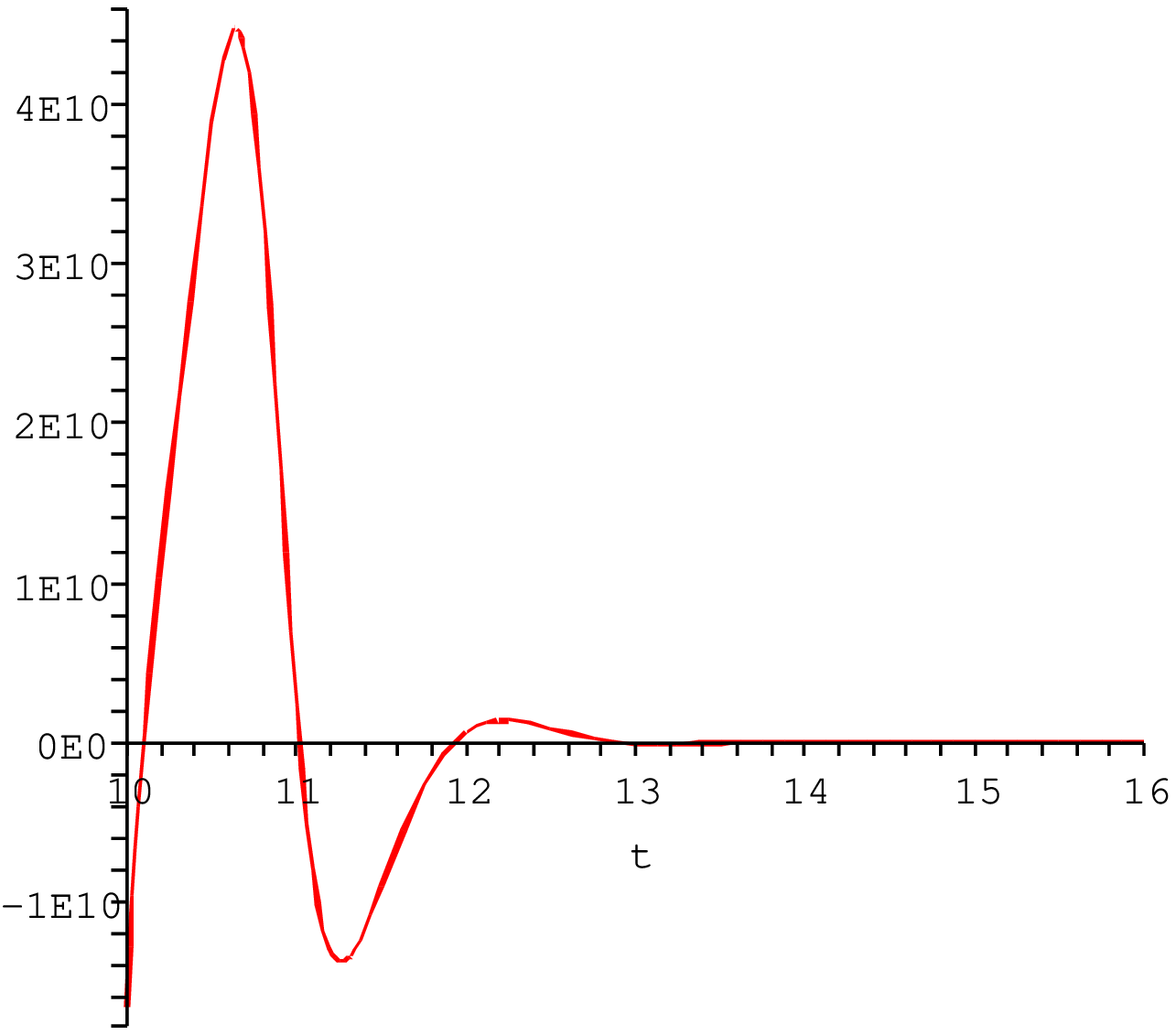}
\includegraphics[width=3.5cm]{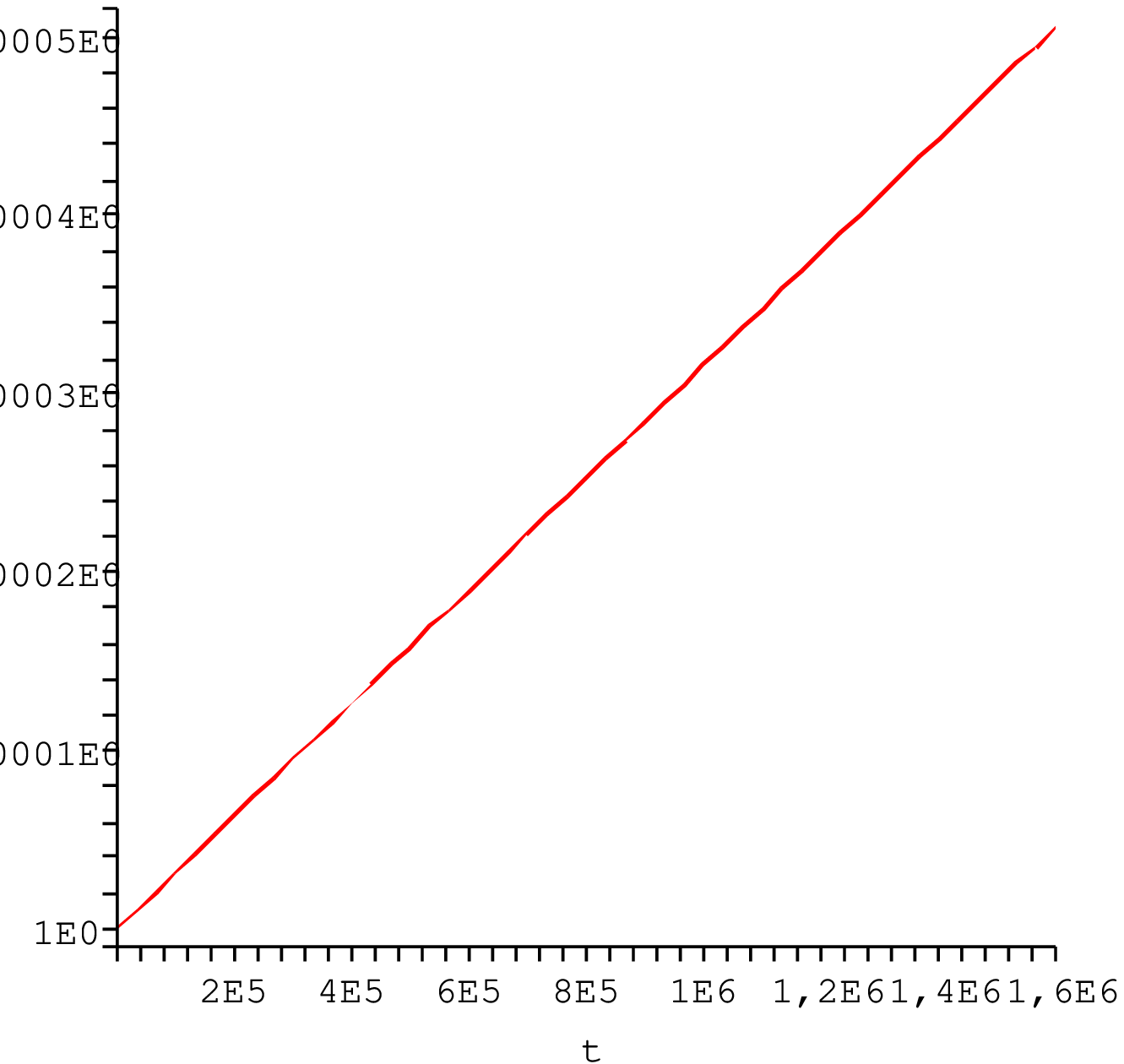}
\caption{From left to right $H$, $w$, $q$, and $a$ for
$\kappa=g_o=\bar{C}=a_0=1$, $\bar{\varphi}=0$ and
$\Lambda_0=3\cdot10^{-21}$.} \label{figure_FRW}}
 We point out that
in spite of the presence of only one scalar field we observe
similarities in the behavior of cosmological quantities with the
two-field model analyzed in \cite{AKVtwofields}. These are crossing
of the phantom divide by $w$ and the qualitative form of plots for
$H$, $w$, and $q$. An order of magnitude of $\Lambda_0$ comes from
an analysis performed in \cite{AKV}.

For $\xi^2=0$ one readily gets using (\ref{solution_dphH_FRW_q20})
for the state and deceleration parameters
\begin{equation}
w=-1-\frac23\frac{2C_H\bar{m}e^{-2\bar{m}t}}{\left(\sqrt{\frac{\Lambda_0}{3}}-
\frac{\kappa}{g_o}C_He^{-2\bar{m}t}\right)^2},\quad
q=-1-\frac{2C_H\bar{m}e^{-2\bar{m}t}}{\left(\sqrt{\frac{\Lambda_0}{3}}-
\frac{\kappa}{g_o}C_He^{-2\bar{m}t}\right)^2}. \label{wq_q20}
\end{equation}
We see that $w$ is always less then $-1$ in this case and a crossing
of the phantom divide does not occur. Thus the tachyon behaves
purely like a phantom for $\xi^2=0$. The scale factor $a$ in this
case is found to be
\begin{equation}
a=a_0e^{\int
Hdt}=a_0\exp\left(\frac{\kappa}{g_o}\sqrt{\frac{\Lambda_0}{3}}t+
\frac{\kappa^2}{g_o^2}\frac{C_H}{2\bar{m}}e^{-2\bar{m}t}\right)\stackrel{t\to\infty}{\approx}
a_0e^{\frac{\kappa}{g_o}\sqrt{\frac{\Lambda_0}{3}}t} \label{a_q20}
\end{equation}
where $a_0$ is an integration constant. Plots for $H$, $w$, $q$, and
$a$ in the case $\xi^2=0$ are presented in
Fig.~\ref{figure_FRW_q20}.
 \FIGURE{
\includegraphics[width=3.5cm]{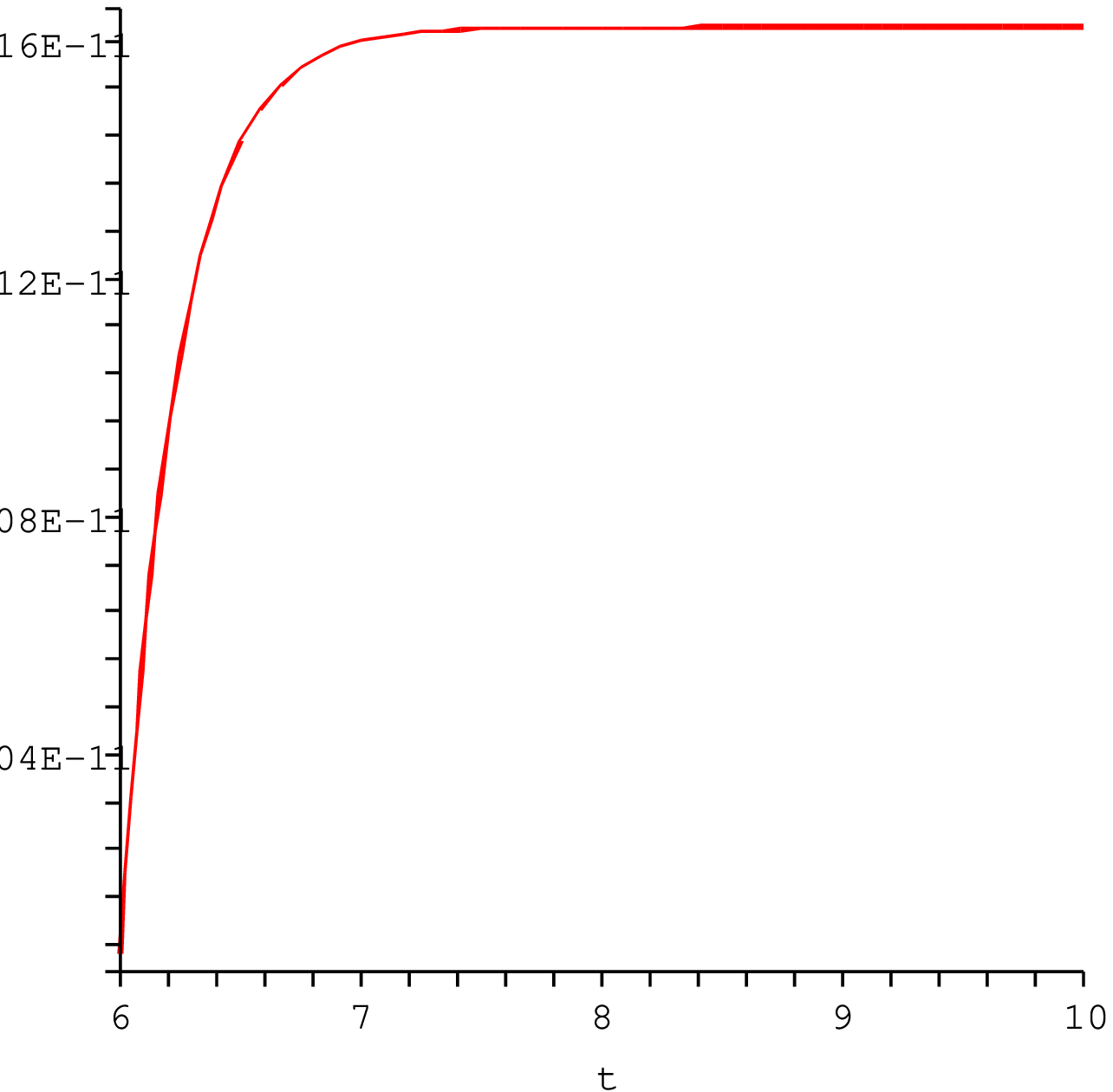}
\includegraphics[width=3.5cm]{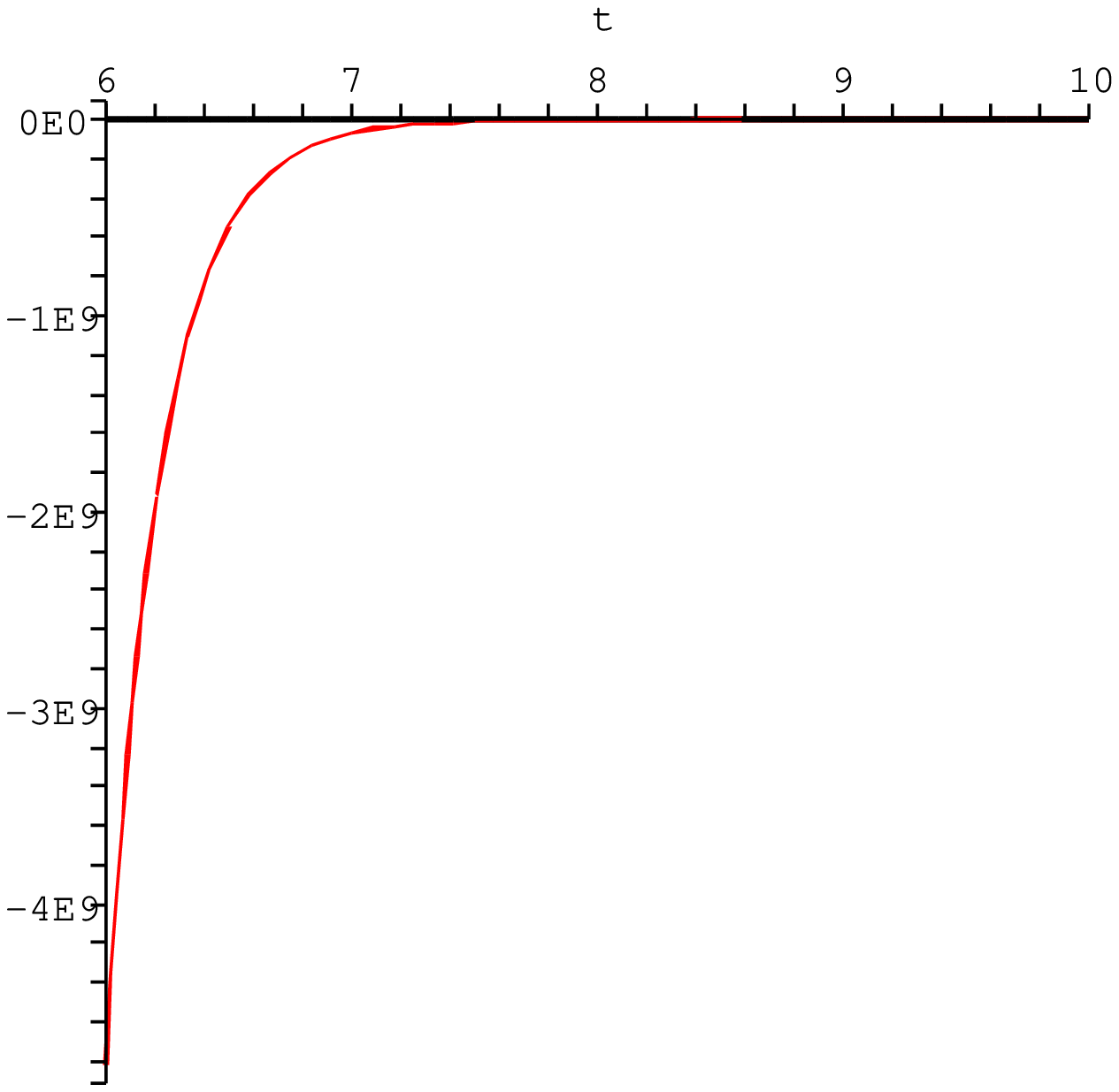}
\includegraphics[width=3.5cm]{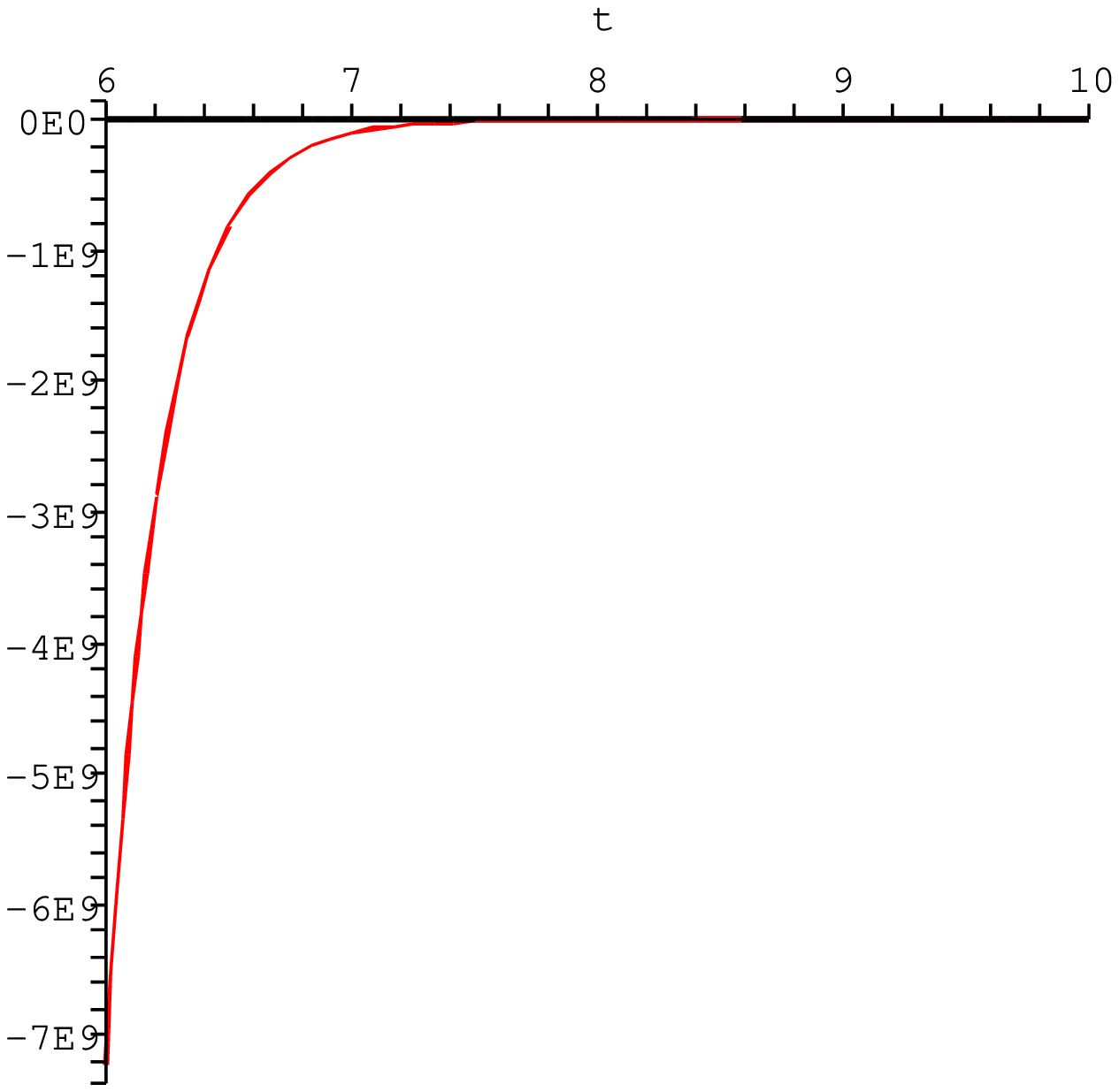}
\includegraphics[width=3.5cm]{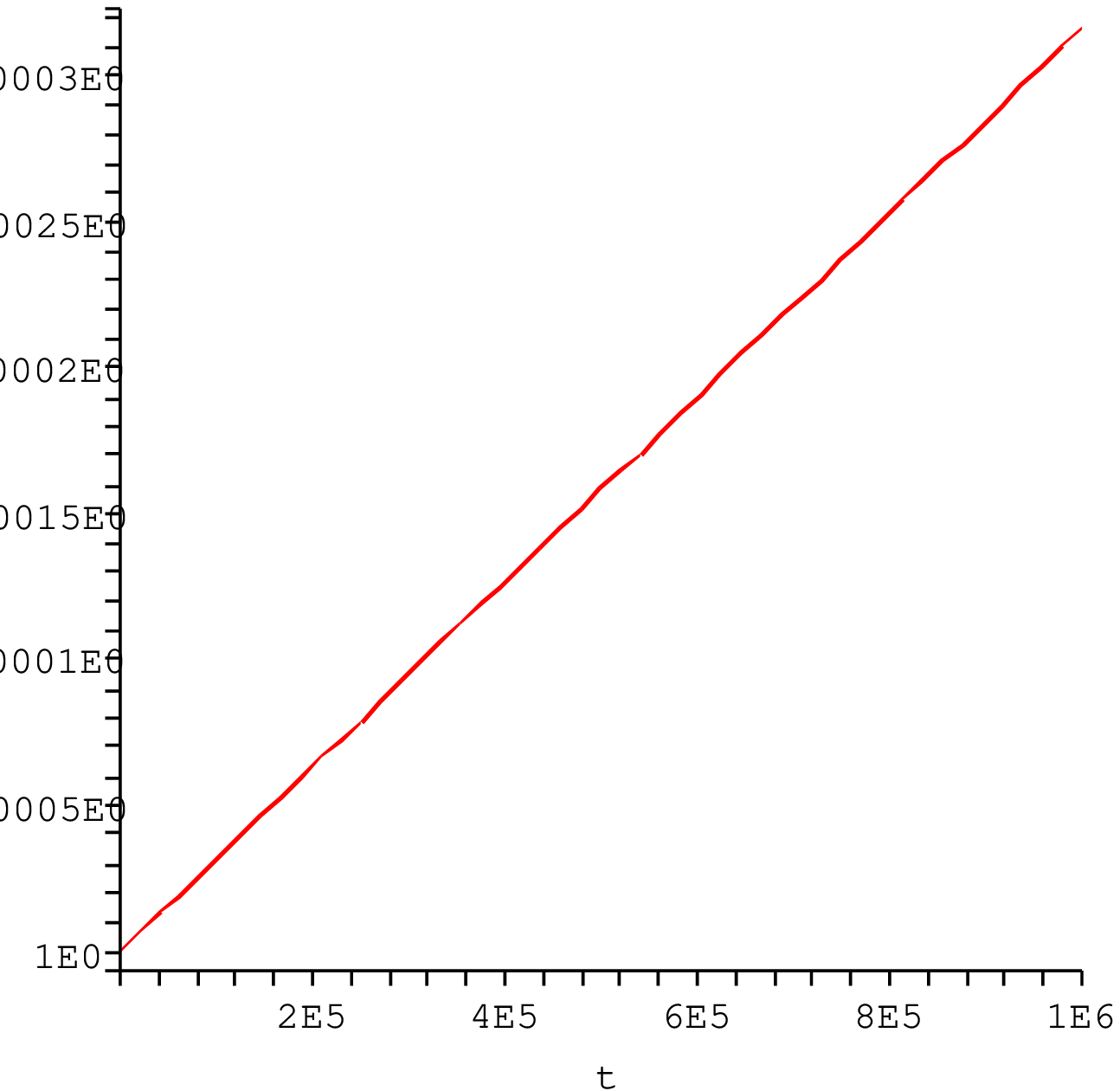}
\caption{From left to right $H$, $w$, $q$, and $a$ for
$\kappa=g_o=\bar{C}=a_0=1$ and $\Lambda_0=3\cdot10^{-21}$ in the
case $\xi^2=0$.} \label{figure_FRW_q20}}
 The behavior of the model
in this case is very close to a single phantom field model solved in
\cite{AKV}.

Remarkably, that both for $\xi^2\approx0.9556$ and $\xi^2=0$ the Big
Rip singularity problem is avoided because $w$ exhibits a
non-trivial time dependence and consequently is not a constant less
then $-1$.

Constructed solution reveals many properties expected from the
cosmological point of view giving rise to an interest to a further
investigation of this model as well as models coming from the SFT in
general. Apart from many other possible directions of developing
this type of models we would like to mention ones which are seen as
the most important.

It would be very interesting and at the same time very difficult to
construct a solution to full non-linear equations
(\ref{EOM_ST0approx}). It seems that if it can be done it will be a
numeric solution. However, even a numeric approach faces a
difficulty of dealing with an exponential of a Beltrami-Laplace
operator in a curved background. The main technical problem is that
a Beltrami-Laplace operator itself contains an unknown function $H$
to be determined while solving the equations.

A complete analysis of a stability of the obtained solution is
also of great importance. A related issue which is an inclusion of
other cosmic fluids and especially the Cold Dark Matter (CDM)
forming about $1/3$ of the present day Universe and an investigation
of a dynamics of such a coupled model and its stability would be
very important (see \cite{AKVCDM} as an example of a coupling of one
phantom field to the CDM).

Also it would be interesting to find an exact solution to full
equations with infinitely many derivatives in curved background
probably by adding extra terms to the action. This is in accord with
an analysis performed in \cite{AKV,AJ} where a small in terms of
coupling constants correction to the potential makes the problem
analytically solvable. An existence of an analytic solution provides
a possibility of a qualitative analysis without a strong support of
numeric methods.

Another important question is a consideration of closed string
scalar fields, the closed string tachyon and dilaton, as well as
their coupling to the open string tachyon. This problem is related
to finding of lump solutions and a development in the flat
background was started in \cite{AJ}. Similar lump solutions have
been constructed in \cite{Trento}. An extension to the FRW Universe
using a linearization of equations of motion developed in the
present paper would be an interesting analysis of the role of closed
string excitations\footnote{See \cite{zw_close} for a discussion on
the closed string tachyon and dilaton condensation.}.

In spirit of a selected role of vector fields in a construction of
local phantom models without an UV pathology \cite{VR} it would be
very interesting to incorporate a string vector field that has in
the SFT approach a non-local interaction in a study of a nonlocal
rolling tachyon dynamics.

%%%%%%%%%%%%%%%%%%%%%%%%%%%%%%%%%%%%%%%%%%%%%%%%%%%%%%%%%%%%%%%%%%%%%%%%%%%%%%%%
%%%%%%%%%%%%%%%%%%%%%%%%%%%%%%%%%%%%%%%%%%%%%%%%%%%%%%%%%%%%%%%%%%%%%%%%%%%%%%%%

\acknowledgments

We are  grateful to L.~Joukovskaya, E.~Kiritsis, T.~Tomaras,
M.~Tsulaia, S.Yu.~Vernov and I.V.~ Volovich for useful discussions.
The work of I.A. and A.K. is supported in part by RFBR grant
05-01-00758, INTAS grant 03-51-6346 and Russian President's grant
NSh-2052.2003.1. The work of A.K. is supported by Marie Curie
Fellowship MIF1-CT-2005-021982 and in part by EU grant
MRTN-CT-2004-512194.

%%%%%%%%%%%%%%%%%%%%%%%%%%%%%%%%%%%%%%%%%%%%%%%%%%%%%%%%%%%%%%%%%%%%%%%%%%%%%%%%
%%%%%%%%%%%%%%%%%%%%%%%%%%%%%%%%%%%%%%%%%%%%%%%%%%%%%%%%%%%%%%%%%%%%%%%%%%%%%%%%

\appendix

\section{A Solution to the characteristic equation}
\label{appendix}

The equation in question is the following
\begin{equation*}
\left(-\xi^2m^2+1\right)e^{\frac{1}{4}m ^2}=3.
\end{equation*}
It is a transcendental one and a solution to it can be represented
analytically as
\begin{equation*}
m^2_k=\frac1{\xi^2}+4W_k\left(-\frac3{4\xi^2}e^{-\frac1{4\xi^2}}\right)=
4\lambda+4W_k\left(-3\lambda e^{-\lambda}\right)% \label{solution_m}
\end{equation*}
where $W_k$ is the $k$-s branch of Lambert $W$ function satisfying a
relation $W(x)e^{W(x)}=x$. It has infinitely many branches
distinguished as $W_k$. A branch analytic at $0$ is referred to as a
principal one. There is exactly one such a branch and $k=0$ is
assigned to it. The branch cut dividing $W_0$ and $W_{\pm1}$ is the
part of the real axis $(-\infty,-1/e)$ and $-1/e$ is the branch
point. The branch cut dividing all other neighbor branches is the
negative real semi-axis and $0$ is the branch point. For the
principal branch $W_0(x)$ the image of the $(-\infty,-1/e)$ interval
is $-\beta\cot\beta+i\beta$, where $\beta\in(0,\pi]$. For the branch
$W_{-1}(x)$ the image of the $(-\infty,-1/e)$ interval is
$-\beta\cot\beta+i\beta$, where $\beta\in(-\pi,0]$. For all other
branches $W_{k}(x)$ the image of the $(-\infty,0)$ interval is
$-\beta\cot\beta+i\beta$, where $\beta\in(2k\pi,(2k+1)\pi]$ if $k>0$
and $\beta\in((2k+1)\pi,(2k+2)\pi]$ if $k<-1$. For our purposes we
are interested in the argument $x=-3\lambda e^{-\lambda}\approx
-0.6042<-1/e$, so the above properties are relevant. To understand
the dependence $-\beta\cot\beta+i\beta$ we are going to consider the
equation
\begin{equation*}
ye^y=x
\end{equation*}
for $x<-1/e$. Assuming a complex solution $y=\alpha+i\beta$ we have
\begin{equation*}
x=(\alpha+i\beta)e^{\alpha+i\beta}=\sqrt{\alpha^2+\beta^2}e^{\alpha}e^{i\left(\arctan\frac{\beta}{\alpha}+\beta\right)}.
\end{equation*}
Since $x$ is a negative real number we have to require a relation
$\arctan\frac{\beta}{\alpha}+\beta=(2k+1)\pi$ which yields
$\alpha=-\beta\cot\beta$ where a branch of the cotangent is the
origin of branches of $W$ function. Using this expression for
$\alpha$ we can write
$x=-\frac{\beta}{\sin\beta}e^{-\beta\cot\beta}$. Here the dependence
on $\beta$ is manifestly odd. From the point of view of $W$ function
the change $\beta\to -\beta$ is the change of branches $k\to -1-k$.
This means the relations
\begin{equation*}
\begin{split}
\jhepre W_k(x)= \jhepre W_{-1-k}(x),\quad \jhepim W_k(x)= -\jhepim
W_{-1-k}(x)
\end{split}
\end{equation*}
hold for $x<-1/e$. Moreover absolute values of real and imaginary
parts of $W_k(x)$ grow when $k$ grows for $k\geq 0$ or $k$ decreases
for $k\leq-1$. The latter is not evident and can be derived from the
series expansion of $W$ function.

For our value of $\xi^2$ all roots are complex with non-zero real
and imaginary parts. Denoting $m^2_k=\alpha_k+i\beta_k$ we have
\begin{equation*}
m_k=r_k+i\nu_k,\quad
r_k=\frac{|\beta_k|}{v},~\nu_k=\sign(\beta_k)\frac12 v,\quad
v=\sqrt{-2\alpha_k+2\sqrt{\alpha_k^2+\beta_k^2}}
\end{equation*}
where for all square roots the arithmetic branch is chosen. In this
case $a_k>0$. Another root is $m_k=-(r_k+i\nu_k)$. Using the
symmetry $\alpha_k=\alpha_{-1-k},~\beta_k=-\beta_{-1-k}$ we have for
our definition of $m_k$ that $m_k=m_{-1-k}^*$. Also for our value of
$\xi^2$ all $\alpha_k<0$.

%%%%%%%%%%%%%%%%%%%%%%%%%%%%%%%%%%%%%%%%%%%%%%%%%%%%%%%%%%%%%%%%%%%%%%%%%%%%%%%%
%%%%%%%%%%%%%%%%%%%%%%%%%%%%%%%%%%%%%%%%%%%%%%%%%%%%%%%%%%%%%%%%%%%%%%%%%%%%%%%%


\begin{thebibliography}{72}

\bibitem{Perlm} S.J. Perlmutter et al., \apj{517}{1999}{565}, [\astroph{9812133}];
A. Riess et al., \asj{116}{1998}{1009}, [\astroph{9805201}].

\bibitem{Riess1} A. Riess et al., \apj{607}{2004}{665}, [\astroph{0402512}].

\bibitem{knop} R.A. Knop et al., \apj{598}{2003}{102}, [\astroph{0309368}].

\bibitem{Tegmark} M. Tegmark et al., \apj{606}{2004}{702}, [\astroph{0310723}].

\bibitem{Spergel} D.N. Spergel et al.,
\newjournal{Astrophys.\ J.\ Suppl.\ }{APJS}{148}{2003}{175}, [\astroph{0302209}].

\bibitem{Spergel06} D.N. Spergel et al., [\astroph{0603449}].

\bibitem{Wetterich} C. Wetterich, \npb{302}{1988}{668}.

\bibitem{Peebles} B. Ratra, P.J.E. Peebles, \apj{325}{1988}{L17}.

\bibitem{Okun} L.B. Okun'. \textit{Leptons and quarks}, Amsterdam, North-Holland,
1982; second edition: Moscow, ``Nauka'', 1990, (in Russian).

\bibitem{S-St}  V. Sahni,  A.A. Starobinsky, \ijmpd{9}{2000}{373},
[\astroph{9904398}].

\bibitem{Padmanabhan-rev} T. Padmanabhan, \prep{380}{2003}{235}, [\hepth{0212290}].

\bibitem{Caldwell} R.R. Caldwell, \plb{545}{2002}{23}, [\astroph{9908168}].

\bibitem{Woodard}
V.K. Onemli, R.P. Woodard, \prd{70}{2004}{107301}, [\grqc{0406098}].

\bibitem{0312430} M. Kaplinghat, S. Bridle, \prd{71}{2005}{123003}, [\astroph{0312430}].

\bibitem{Riess06} A.G. Riess and M. Livio,
[\astroph{0601319}].

\bibitem{sami_review}
E. J. Copeland, M. Sami, Sh. Tsujikawa, [\hepth{0603057}].

\bibitem{modGR}
G.Allemandi, A. Borowiec, M. Francaviglia, \prd{70}{2004}{103503},
[\hepth{0407090}]; I. P. Neupane, [\hepth{0602097}]; Sh. Nojiri, S.
Odintsov, M. Sami, [\hepth{0605039O}].

\bibitem{Arefeva} I.Ya. Aref'eva, [\astroph{0410443}].

\bibitem{NPB} I.Ya.~Arefeva, D.M.~Belov, A.S.~Koshelev,
P.B.~Medvedev, \npb{638}{2002}{3}, [\hepth{0011117}].

\bibitem{Sen-g} A.~Sen, [\hepth{0410103}].

\bibitem{brane}
 G.~Dvali, G.~Gabadadze and M.Porrati, \plb{485}{2000}{208}, [\hepth{0005016}];
 V.~Sahni, Y.~Shtanov, \newjournal{J.\ Cosm.\ and\ Astropart.\ Phys.\ }{JCAPA}{0311}{2003}{014}, [\astroph{0202346}];
 R.~Kallosh, A.Linde, \prd{67}{2003}{023510}, [\hepth{0208157}];
 Sh.~Mukohyama, L.~Randall, \prl{92}{2004}{211302}, [\hepth{0306108}];
 Ph.~Brax, C.~van de Bruck, A.-C.~Davis, \newjournal{Rept.\ Prog.\ Phys.\ }{67}{2004}{2183}, [\hepth{0404011}];
 Th.N.~Tomaras, [\hepth{0404142}];
 E.J.~Copeland, M.R.~Garousi, M.~Sami, Sh.~Tsujikawa, \prd{71}{2005}{043003}, [\hepth{0411192}];
 G.~Kofinas, G.~Panotopoulos, T.N.~Tomaras, \jhep{0601}{2006}{107}, [\hepth{0510207}];
 R.-G.~Cai, Y.~Gong, B.~Wang,  \newjournal{J.\ Cosm.\ and\ Astropart.\ Phys.\ }{JCAPA}{0603}{2006}{006}\textbf{0603},
[\hepth{0511301}];
 P.S.~Apostolopoulos, N.~Tetradis, [\hepth{0604014}].


\bibitem{review-sft} K.~Ohmori,  [\hepth{0102085}];
 I.Ya.~Aref'eva, D.M.~Belov, A.A.~Giryavets, A.S.~Koshelev,
P.B.~Medvedev, [\hepth{0111208}];
 W. Taylor,  [\hepth{0301094}].

\bibitem{SD} Al. Adams, N. Arkani-Hamed, S. Dubovsky, Al.
    Nicolis, R. Rattazzi, [\hepth{0602178}].

\bibitem{VR} V.A. Rubakov, [\hepth{0604153}].

 \bibitem{AJK} I.Ya.~Aref'eva, L.V.~Joukovskaya, A.S.~Koshelev,
 \jhep{0309}{2003}{012}, [\hepth{0301137}].

\bibitem{yar} Ya.I. Volovich, \jpha{36}{2003}{8685},  [{\tt math-ph/0301028}].

\bibitem{VV} V.S. Vladimirov, Ya.I. Volovich, \tmp{138}{2004}{297}, [{\tt math-ph/0306018}].

\bibitem{Zw} N. Moeller, B. Zwiebach, \jhep{0210}{2002}{034},
[\hepth{0207107}].

\bibitem{VSV} V.S.~Vladimirov, \newjournal{Izvestya\ RAN\ }{IZRAN}{69}{2005}{55}, (in Russian).


\bibitem{calcagni} G. Calcagni, \jhep{05}{2006}{012},
[\hepth{0512259}].

%%%%%%%%%%%%%%%%%%%%%%%%%%

\bibitem{AKVtwofields} I.Ya. Aref'eva, A.S. Koshelev, S.Yu. Vernov, \prd{72}{2005}{064017}, [\astroph{0507067}].

\bibitem{Bo}
Bo Feng, Mingzhe Li, Yun-Song Piao, Xinmin Zhang,
[\astroph{0407432}].

\bibitem{Wei} H. Wei, R.-G. Cai, \plb{634}{2006}{9}, [\astroph{0512018}].

\bibitem{mukhanov}
C. Armendariz-Picon, V. Mukhanov, P.J. Steinhardt,
\prl{85}{2000}{4438}, [\astroph{0004134}];
 C. Armendariz-Picon, V. Mukhanov, P.J. Steinhardt, \prd{63}{2001}{103510}, [\astroph{0006373}].


\bibitem{Vikman}A. Vikman, \prd{71}{2005}{023515}, [\astroph{0407107}].

\bibitem{andrianov}
A.A. Andrianov, F. Cannata, A.Y. Kamenshchik,
\prd{72}{2005}{043531}, [\grqc{0505087}].


\bibitem{AMZ} I.Ya.~Aref'eva, P.B.~Medvedev, A.P.~Zubarev, \npb{341}{1990}{464}.

\bibitem{PTY} C.R.~Preitschopf, C.B.~Thorn, S.A.~Yost, \npb{337}{1990}{363}.

\bibitem{BFOW} L.~Brekke, P.G.O.~Freund, M.~Olson, E.~Witten,
\npb{302}{1988}{365}.

\bibitem{padic} V.S.~Vladimirov, I.V.~Volovich, E.I.~Zelenov,
\textit{p-adic Analysis and Mathe\-matical Physics}, WSP, Singapore,
1994.


\bibitem{AKV} I.Ya.~Aref'eva,
A.S.~Koshelev, S.Yu.~Vernov,  [\astroph{0412619}].

\bibitem{AKVCDM} I.Ya.~Aref'eva,
A.S.~Koshelev, S.Yu.~Vernov,  \plb{628}{2005}{1},
[\astroph{0505605}].

\bibitem{AJ} I.Ya. Aref'eva and L.V. Joukovskaya, \jhep{0510}{2005}{087},
[\hepth{0504200}].

\bibitem{Trento}  V. Forini, G. Grignani, G. Nardelli,
[\hepth{0502151}].


\bibitem{zw_close} H. Yang, B. Zwiebach,
\jhep{0508}{2005}{046}, [\hepth{0506076}];
 H. Yang, B. Zwiebach, \jhep{0509}{2005}{054}, [\hepth{0506077}].

\end{thebibliography}
\end{document}